%% file: main.tex
\documentclass[twocolumn]{aastex631}
\pdfoutput=1 
\usepackage[T1]{fontenc}
\usepackage[utf8]{inputenc}
\usepackage[english]{babel}

\usepackage{microtype}  
\usepackage{amsmath}
\usepackage{amsfonts}
\usepackage{amssymb}
\usepackage{multirow}
\usepackage{tikz}
\usepackage{xcolor}
\usepackage{soul}

\definecolor{pink}{RGB}{232,132,161}
\definecolor{yellow}{RGB}{255,213,0}

\newcommand{\affuofa}{University of Arizona, 933 N. Cherry Ave,
    Tucson, AZ 85721, USA}
\newcommand{\affuofu}{Department of Physics and Astronomy, University of Utah, 115 South 1400 East, Salt Lake City, Utah 84112, USA}
\newcommand{\affuva}{Department of Astronomy, University of Virginia, 530 McCormick Road, Charlottesville, VA 22904, USA}
\newcommand{\affunam}{Instituto de Radioastronom\'ia y Astrof\'isica, Universidad Nacional Aut\'onoma de M\'exico, Apdo. Postal 72-3, 58089 Morelia, Mexico}

\newcommand{\chambe}{\citet{Chamberlain2024}}

\sloppypar

\input{math_definitions.tex}

\begin{document}

\title{A Physically Motivated Framework to Compare Merger Timescales\\ of Isolated Low- and High-Mass Galaxy Pairs Across Cosmic Time}
\shorttitle{Merger Timescales in TNG100}

\author[0000-0001-8765-8670]{Katie~Chamberlain}
\affiliation{\affuofa}

\author[0000-0002-9820-1219]{Ekta~Patel}
\thanks{NASA Hubble Fellow}\affiliation{\affuofu}

\author[0000-0003-0715-2173]{Gurtina Besla}
\affiliation{\affuofa}

\author[0000-0002-5653-0786]{Paul Torrey}
\affiliation{\affuva}

\author[0000-0002-9495-0079]{Vicente Rodriguez-Gomez}
\affiliation{\affunam}

\shortauthors{Chamberlain et al.}

\begin{abstract}
    The merger timescales of isolated low-mass pairs ($\rm 10^8<M_*<5\times10^9\,\Msun$) on cosmologically motivated orbits have not yet been studied in detail, though isolated high-mass pairs ($\rm 5\times10^9<M_*<10^{11}\,\Msun$) have been studied extensively.
    It is common to apply the same separation criteria and expected merger timescales of high-mass pairs to low-mass systems, however, it is unclear if their merger timescales are similar, or
    if they evolve similarly with redshift. 
    We use the Illustris TNG100 simulation to quantify the merger timescales of isolated low-mass and high-mass major pairs as a function of cosmic time, and explore how different selection criteria impact the mass and redshift dependence of merger timescales. 
    In particular, we present a physically-motivated framework for selecting pairs via a scaled separation criteria, wherein pair separations are scaled by the virial radius of the primary's FoF group halo ($r_{\mathrm{sep}}< 1 \Rvir$).  
    Applying these scaled separation criteria yields equivalent merger timescales for both mass scales at all redshifts.
    Alternatively, static physical separation selections applied equivalently to all galaxy pairs at all redshifts leads to a difference in merger rates of up to $\sim1\Gyr$ between low- and high-mass pairs, particularly for $\rsep<150\kpc$. 
    As a result, applying the same merger timescales to physical separation-selected pairs will lead to a bias that systematically over-predicts low-mass galaxy merger rates.

\end{abstract}

\section{Introduction} \label{sec:intro}
Galaxy merger rates provide an important pathway for tests of hierarchical assembly from $\Lambda$CDM theory, and are critical for understanding the formation and evolution of galaxies across time~\citep[e.g.][]{Stewart2009, Hopkins2010b,RG2015}. 
In the era of future surveys with JWST, Rubin, and Roman, these tests will be extended to higher redshifts and lower mass scales than were previously accessible~\citep{Gardner2006,Spergel2015,Robertson2019a,Robertson2019b,Behroozi2020,Martin2022}.  

At higher redshift ($z\gtrsim1$) and at lower masses ($\ms{} \lesssim10^{9.5}$), even isolated galaxies are highly disturbed~\citep{Wuyts2012,Wuyts2013,Martin2018,Martin2021,Varma2022}, making close pair fractions a critical alternative to morphological signatures for merger rate studies.
However, there is currently no framework for comparing merger rates for low-mass and high-mass galaxies as a function of time, and tensions between predictions from theory and observational studies exist in the literature at low masses \citep{RG2015}. 

This study seeks to establish a framework to enable comparisons of merger rates across mass scales and redshift so that merger rates of low-mass galaxies can be used as tests of $\Lambda$CDM theory. 
These results can also be used to interpret observations of galaxy morphologies and their hierarchical evolution, including the role of mergers in triggering starbursts, fueling active galactic nuclei (AGN), facilitating the formation of tidal features, etc., in low-mass galaxies~\citep[e.g.][]{Stierwalt2015,Pearson2016,Privon2017,Kristensen2021,Martin2021,Luber2022,Martin2022,GuzmanOrtega2023,ByrneMamahit2024,KadoFong2024}. 

In our previous work, we show that the redshift evolution of pair fractions of isolated low-mass and high-mass pairs in Illustris TNG100 differ significantly, particularly at $z<3$ where pair fractions of low-mass pairs decrease up to $75\%$ between $z=3$ and $z=0$, while high-mass pair fractions peak at $z=0$~\citep{Chamberlain2024}. 
We suggested that differences in the merger timescales of low-mass and high-mass pairs might be the cause of the pair fraction evolution differences.
In our following analysis, we will investigate this question directly and answer whether merger timescales are responsible for these pair fraction differences.

Additionally, we found in \citet{Chamberlain2024} that recovering the pair fraction differences seen between the two mass scales is sensitive to the selection criteria adopted during pair selection.
Physical separation cuts applied equivalently to all pairs, with no mass or redshift dependence, eliminate the ability to distinguish the behavior of the pair fraction evolution of low-mass and high-mass pairs, even for separation cuts as large as $\rsep<300\kpc$. 
Alternatively, when separation criteria vary with the mass and redshift of each system, in particular when the separation is scaled by the virial radius $\Rvir$ of the pair's FoF group halo, the ability to distinguish the underlying pair fraction behavior was recovered for scaled separations as low as $\rsep < 0.5\,\Rvir$. 
Employing scaled separation criteria then permits the equivalent comparison of pair fractions between low-mass and high-mass pairs, and robust comparisons across redshifts from $z=0-4$. 
Since merger rate estimates are derived from close pair fractions and merger timescales, the results from our previous work imply that careful consideration of separation criteria is required for merger timescale and merger rate studies at different mass scales and redshifts as well.

In order to investigate the impact of pair selection criteria on merger timescales specifically, we will track the pairs selected in \chambe{} forwards and backwards in time to construct orbits for each pair.
We will then study the merger timescales of isolated low-mass and high-mass pairs for the same set of physical and scaled separation criteria used in \chambe{} to determine whether merger timescales of isolated pairs are robust to the selection criteria used to determine the pair samples.

In this paper, we aim to extend the framework for pair selection criteria presented in \chambe{}.
In Sec.~\ref{sec:methods}, we detail our selection criteria for isolated low-mass and high-mass orbits in the Illustris TNG100 simulation. 
We examine the redshift evolution of the number of pairs and merger fraction of our sample in Sec.~\ref{sec:pairprops}.
In Sec.~\ref{sec:results}, we present our findings on the separation and redshift dependence of the merger timescales of low-mass and high-mass pairs, and study the impact of physical and scaled separation criteria for pair selection. 
Finally, we discuss the implications of this work and provide suggestions for a self-consistent way of studying pairs across redshifts and mass scales in Sec.~\ref{sec:discussion}, and present our final conclusions in Sec.~\ref{sec:conclusions}.

\section{Methodology} \label{sec:methods}
The IllustrisTNG simulation suite~\citep{TNG1,TNG2,TNG3,TNG4,TNG5} is a set of large volume dark-matter-only and full magnetohydrodynamical cosmological simulations consistent with the \textit{Planck 2015} \lcdm{} cosmology~\citep{Planck2015}. 

Following~\chambe{}, we use TNG100-~1 (hereafter TNG100), which is the highest resolution full physics run of the ($100\Mpc)^3$ volume.
In particular, we utilize the group catalogs produced by the \subfind{} algorithm~\citep{Springel2001b,Dolag2009} and the merger tree catalogs generated by the \sublink{} algorithm~\citep{RG2015}.
The catalogs consist of a set of 100 snapshots, ranging from $z\sim20$ (snapshot 0) to $z=0$ (snapshot 99).

Our sample consists of major low-mass and high-mass pairs (stellar mass ratio $1/4 \leq M_{*2}/M_{*1}\leq 1$) that are isolated, but physically associated, as in \citet{Chamberlain2024}.
From this sample, we will determine the fraction of pairs at each snapshot that merge before $z=0$, and track the orbits of each pair to study their merger timescales from $z=0-6$.

\subsection{Pair sample}
We begin with an extension of the \paircat{} described in \chambe{}, which consists of a collection of isolated galaxy pairs at each snapshot in TNG100. 
We collect our base sample at each redshift in the simulation. 
A brief version of the selection routine is transcribed here for completeness, and we refer readers to Sec.~2 of \chambe{} for more detail and a discussion of the selection criteria choices. 

At each snapshot, low-mass and high-mass pairs are chosen by first selecting the two most massive subhalos (by assigned stellar mass, using abundance matching; as described later in this subsection) from FoF groups with virial mass $\MG$:
\begin{align*}
        \mbox{\textbf{low mass:}}&\,\rm \MG = 8\times 10^{10}- 5\times 10^{11}\,\Msun \\ 
        \mbox{\textbf{high mass:}}&\, \rm \MG =1\times 10^{12}- 6.5\times10^{12}\,\Msun.
\end{align*}
Virial masses (and radii) are calculated according to the spherical collapse model of \citet{Bryan1998}, and are provided by the TNG group catalogs.
Requiring pairs to belong to the same FoF group ensures that the pairs are distant from other massive nearby systems that could perturb the dynamical state of the pair.  
In addition, for each pair, we calculated the Hill Radius of every FoF group with a higher mass within $10\,\Mpc$ of the pair, as a proxy to discern whether the pair lies outside the gravitational sphere of influence of these more massive groups.
We found that over 99\% of our pairs at $z=0$ are more distant than two times the Hill Radius of each FoF group.

We require the subhalos that constitute a pair to meet a minimum subhalo mass, \Mhalo, criteria of 
\begin{equation*}
    \mbox{\textbf{minimum subhalo mass:}}\,
    \Mhalo > 1\times10^{9}\Msun.
\end{equation*}
at the snapshot of consideration. 
For TNG100, this ensures that subhalos are resolved into over 
$\sim$100 particles, enough to robustly identify gravitationally bound subhalos in the \subfind{} and \sublink{} catalogs.
For each subhalo in the FoF group that passes the minimum subhalo mass criteria, we utilize the \sublink{} catalogs to find the peak halo mass of each subhalo \citep{RG2015}. 

As in \chambe{}, stellar masses are assigned to each subhalo in the FoF group using the abundance matching prescription of \citet{Moster2013}. 
Utilizing abundance matching to assign stellar masses allows us to circumvent simulation-specific stellar mass effects, and also results in a more straightforward process for applying our methodology to observational studies (see Sec.~\ref{sec:disc-suggestions} for more details). 
The peak halo mass and current redshift are used to calculate the stellar mass of each subhalo via the abundance matching prescription  $\ms{}=f(\Mpeak,z)$, where $z$ is the redshift of consideration.

In \chambe{}, the
abundance matching prescription was sampled 1000 times for each subhalo to account for the spread in the Stellar Mass -- Halo Mass (SMHM) relation. 
For the present study, we only use the stellar masses given by the median of the abundance matching relationship. 
We expect that the spread of merger timescales from different orbital configurations, as well as the redshift spacing of the TNG100 snapshots, will dominate over uncertainties from the number of pairs in the catalog, which varied by $\sim3\%$ \citep[see][]{Chamberlain2024}.

Primary subhalos are defined as the subhalo with the highest assigned stellar mass, $M_{*1}$ in the FoF group, and secondaries are defined as the second most massive subhalo with stellar mass $M_{*2}$. 
Our sample of major pairs then consists of all pairs of primary and secondary subhalos with 
\begin{align*} 
\mbox{\textbf{low mass primaries:}}&\, 10^{8}< M_{*1} < 5\times10^{9} \Msun \\ 
\mbox{\textbf{high mass primaries:}}&\, 5\times 10^{9}< M_{*1} < 10^{11} \Msun\\
\mbox{\textbf{stellar mass ratio:}}&\,      
    M_{*2}/M_{*1} > 1/4.
\end{align*}
A primary or secondary subhalo can only be a member of one single pair at a given snapshot, such that a collection of $N$ pairs consists of $N$ unique primaries and $N$ unique secondaries.
A subhalo can belong to multiple different pairs at different redshifts. 
For example, the primary of a pair that merges at $z=2$ can be selected with a different secondary at $z=1$, constituting a new pair. 
More detail regarding the uniqueness of pairs and orbits is discussed in Sec.~\ref{sec:methods-unique}.


The base sample of pairs used for this analysis then consists of the set of all
isolated 
low-mass and high-mass major pairs from each redshift of the TNG100 simulation. 

\subsection{Mergers} \label{subsec:mergers}
Here we describe how mergers are identified. 
Note that this definition is specific to subhalo mergers and therefore results are not guaranteed to hold for galaxy mergers~\citep[see e.g.,][]{RG2015,Patton2024}, especially in cases where dark matter subhalos and galaxies have different centers of mass, as is the case in our own Galaxy~\citep{Gomez2015,GaravitoCamargo2019,Petersen2021,Chamberlain2023}. 

The primary and secondary subhalos of each pair have a \texttt{SubfindID} in the \subfind{} merger tree catalogs, which contain information about every FoF group and subhalo at each redshift. 
The \sublink{} catalogs track subhalos from one redshift to the next, and thus enable us to track subhalos both backwards and forwards in time from any given redshift. 

We utilize the \texttt{DescendantID} field of the \sublink{} catalogs to determine which pairs from our base sample merge before the end of the simulation (at $z=0$) and when each merger occurs. 
The \texttt{DescendantID} field provides the \texttt{SubhaloID}\footnote{Note that the \texttt{SubhaloID} is distinct from the \texttt{SubfindID}, and is unique for every subhalo in the merger trees. 
The \sublink{} catalogs provide the associated \texttt{SubfindID} of each subhalo.} of the subhalo's descendent in the next (or one of the following) snapshots, if it has one. 

In this analysis, a pair is classified as a merger if the primary and secondary subhalo share the same \texttt{DescendantID} at the same redshift. 
This means that the primary and secondary subhalo have merged such that \subfind{} can no longer distinguish them as two separate halos, therefore yielding a single descendant subhalo.
If a primary and secondary never share the same \texttt{DescendantID} at the same redshift, the pair is defined as a `non-merger.'

For each merging pair, we define the \textbf{merger redshift} as the redshift that immediately follows the first snapshot where the primary and secondary have the same \texttt{DescendantID}. 
For example, if the primary and secondary have different \texttt{DescendantIDs} from $z=6$ to $z=2$, but have the same \texttt{DescendantID} at $z=2$, then the merger must take place between $z=2$ and $z=1.9$, which is the next redshift corresponding to a snapshot in the simulation. 
In this example, we take $z=1.9$ to be the merger redshift. 

\subsection{Orbits} 
\begin{figure}[tb]
    \begin{center}
    \includegraphics[width=\columnwidth]{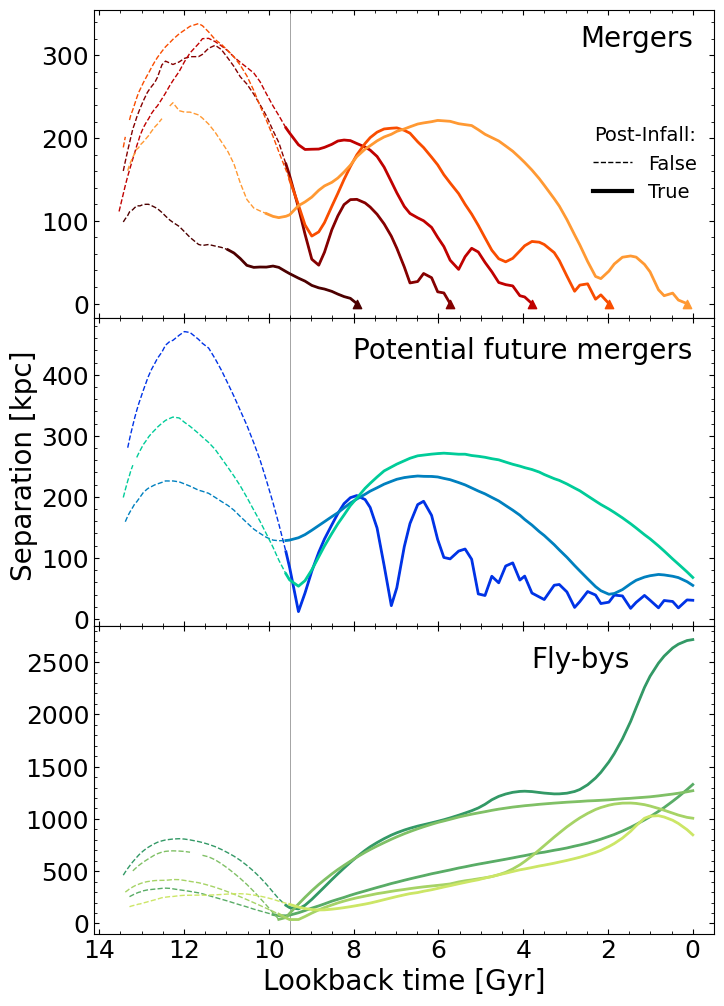}
    \caption{Selection of example orbits of low-mass major pairs that pass the pair selection criteria at $z=1.5$, showing the separation between the primary and secondary as a function of lookback time. 
    (Top) Orbits of pairs that merge before $z=0$.
    (Middle) Orbits of pairs that do not merge before $z=0$ (non-mergers), but are likely to merge if the simulation continued. 
    (Bottom) Orbits of pairs that do not merge before $z=0$ (non-mergers) and are unlikely to do so in the future $\sim2\Gyr$ (past the end of the simulation).
    Solid lines represent the post-infall orbits, i.e. after the pair share a common FoF group.
    Dashed lines show the pre-infall portion of the orbit. 
    Triangle points in the first panel show the first redshift after merger where the orbit has a separation of $0\,\kpc$.
    The vertical grey line marks $z=1.5$ at a lookback time of $9.5\,\Gyr$. 
    }
    \label{fig:example-orbits}
    \end{center}
\end{figure}

We extract orbits for all mergers and non-mergers in our base pair sample using the merger tree catalogs. 
An orbit for a single pair is defined to be the physical separation between the primary and secondary subhalo as a function of redshift (or lookback time).

A given pair from the base sample at redshift $z_n$ passes all of the selection criteria from \chambe{} at $z_n$, and can additionally be followed backwards and forwards in time using the \sublink{} merger trees. 
We track the positions of both the primary and secondary subhalo at each redshift and calculate the physical separations after accounting for the periodic boundary conditions of the simulation box.
In cases where the primary or secondary does not have a defined position at a given redshift, we do not compute a value for the separation at that redshift.\footnote{If a subhalo is very small, or is passing through a more massive subhalo, and is unable to reach the density contrast required to be identified as an independent structure by the \subfind{} algorithm, it will not have a defined position in the \sublink{} catalogs. The \sublink{} algorithm allows for subhalos to skip a single snapshot, and identifies the `skipped descendent' in the $S_{n+2}$ snapshot, so that the orbit can be evaluated before and after the skip occurs. See Sec.~3 in ~\citet{RG2015} for more details.}

In addition to the physical separation, we also calculate the scaled separation of a pair at each redshift. We use the definition of scaled separation \begin{equation}\label{eq:scaled}
r_{\rm sc} \equiv\rsep/\Rvir
\end{equation}

from~\cite{Chamberlain2024}, where $\rsep$ is the physical separation in $\kpc$, and $\Rvir$ is the virial radius of the pair's FoF group.\footnote{Given by \texttt{Group\_R\_TopHat200} in the group catalogs. If the secondary is not in the same FoF group as the primary, the virial radius used to calculate the scaled separation will remain that of the primary's FoF group, which merging secondaries will eventually re-enter prior to merger.} 
The virial radius of the pair's FoF group reasonably approximates the virial radius of a halo with a virial mass equal to the combined subhalo mass of the primary and secondary.
This ``scaled separation" is, by construction, a function of mass and redshift.


Note that we choose not to interpolate the orbits to get a more precise determination of the merger timescale, as the spread on the merger timescales in our study will be dominated by the variation of orbital configurations, rather than the uncertainty due to snapshot spacing. However,~\citet{Patton2024} recently showed that for studies needing more precise determination of, e.g., pericenter passages or merger times, their 6D interpolation scheme can improve timing uncertainties from $\pm80\Myr$ to $\pm3.3\Myr$.

\subsubsection{Defining Post-Infall}
\label{subsubsec:post-infall}
While the orbit of a pair may be calculated at very early times, we wish to constrain our orbital analysis to only physically associated pairs, and thus will not consider the orbit of a subhalo pair before they belong to the same FoF group. 
Specifically, we will consider only the ``post-infall" part of each pair's orbit. 

We define the redshift of ``first infall" as the redshift of the first snapshot where the primary and secondary have the same parent FoF halo, and post-infall as all following redshifts starting with the redshift of first infall. We define \textbf{merger timescales} to begin at the time of first infall and conclude at the time of merger redshift, as defined in Section \ref{subsec:mergers}.

Figure~\ref{fig:example-orbits} presents a set of example orbits, which shows the wide variety of orbit-types that can be found in the TNG100 simulation for pairs that were originally selected from \chambe{} at $z=1.5$. 
The top panel shows the orbits of five pairs that merge, where solid lines show the post-infall parts of the orbit, and dashed lines show the pre-infall portions of the orbit before the secondary and primary ever share a FoF group halo. 
Even amongst galaxies of the same approximate stellar mass and stellar mass ratio, the spread of merger timescales can be very large, with merger timescales between $\sim1-10\,\Gyr$ from first infall to merger (triangles). 

For some pairs, the secondary subhalos can experience first infall, then after one pericentric passage can return to large distances where they are temporarily assigned a different FoF group halo than that of their primary. 
In these cases, we use the full orbit beginning at first infall, including the segments of the orbit where the secondary is assigned to another FoF group temporarily,  through to merger as described in Section \ref{subsec:mergers}.
This definition is most robust for considering the full interaction timescales of merging pairs.

The middle panel of Fig.~\ref{fig:example-orbits} (``Potential Future Mergers") shows the orbits for three pairs that are classified as non-mergers, since they do not merge before $z=0$, but which appear likely to merge within a few Gyr past the end of the simulation. 
These orbits can have a variety of orbital periods, and the number of pericentric passages can vary significantly. 
The two lighter blue orbits are very long period orbits with 1-2 pericenter passages in the past $10\,\Gyr$, while the darker blue orbit has a much shorter period with three close passages in just the past $2\,\Gyr$.

The bottom panel shows the orbits of ``Fly-by" interactions (non-mergers) that are unlikely to merge in the near future, if ever.  
Note that we do not split our non-merger category into fly-bys and potential future mergers for any of our following analyses, and such distinctions were made only for the purposes of showing the diversity of selected orbits. We further discuss the role of non-mergers and fly-bys on our results in Sec.~\ref{sec:disc-nonmergers}.

\subsubsection{Uniqueness of orbits}
\label{sec:methods-unique}
Since we collect the orbit for each pair at all redshifts after infall, a singular orbit will be selected as many times as the number of redshifts (or snapshots) where that pair exists. However, it is only necessary to keep a single instance of any given orbit in our catalog to avoid skewing our data artificially to longer merger timescales.

To distinguish the collection of unique orbits, each pair is assigned a `pairkey' while constructing their orbit. 
The pairkey is created by concatenating the earliest \texttt{SubhaloID} of the primary and secondary subhalo from the \sublink{} catalogs, and is unique for each pair of halos. 
After each pair is assigned a unique pairkey, we only keep one instance of an orbit per pair to avoid double/multi-counting in our orbit catalog.

Note that a single subhalo may be a member of many different pairs, but will only have one unique orbit per unique pair.
For example, the primary of a low-mass pair selected at $z=3$ that merges before $z=2$ may be selected via our selection criteria again at $z=1$ with a new secondary companion. 
In this case, both orbits (of the original low-mass pair and the new pair which includes the primary subhalo from the previous merger) are retained in the orbit catalog.

The total number of orbits, before removing the redundant orbits,  
is 71,429 for low-mass pairs, and 20,824 high-mass pairs. 
However, after removing all redundant orbits, there remain 22,213 low-mass orbits, and 3,029 high-mass orbits that each correspond to a unique pair of subhalos. The collection of all unique orbits constitutes our orbit sample and is the dataset that will be used for the remainder of this analysis.

\section{Pair Sample Properties}\label{sec:pairprops}
\begin{figure}[tb]
    \begin{center}
    \includegraphics[width=\columnwidth]{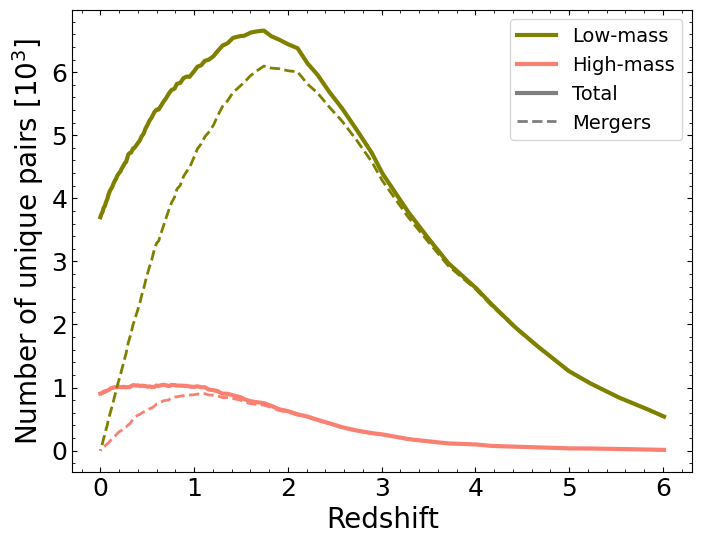}
    \caption{The number of unique low-mass (green) and high-mass (pink) pairs in our orbit catalog as a function of redshift. Orbits are only defined between first infall and the merger redshift (for merging pairs) or $z=0$ (for non-merging pairs). 
    The solid lines show the total number of pairs at a given redshift (including non-mergers and mergers), while dashed lines show the number of pairs at a given redshift that will merge before $z=0$.
    The decrease in the number of unique low-mass pairs from $z=2$ to $z=0$ is due to pairs being removed from the sample via mergers.
    }
    \label{fig:numorbits}
    \end{center}
\end{figure}

\begin{figure}[htb]
    \begin{center}
    \includegraphics[width=\columnwidth]{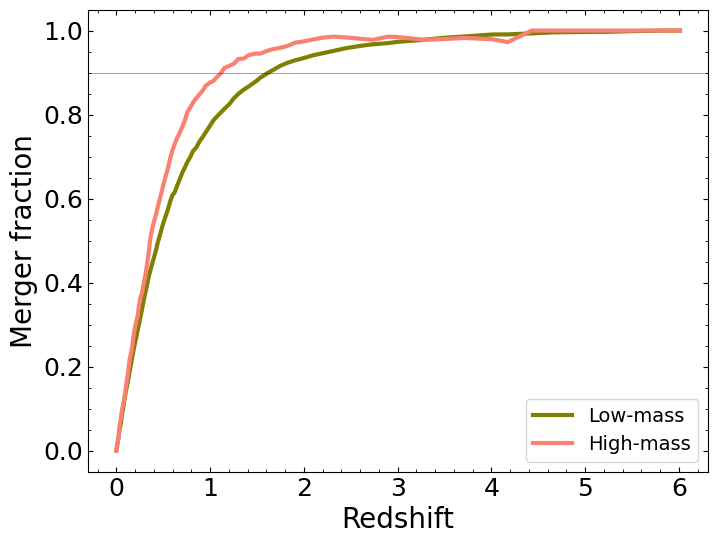}
    \caption{The fraction of low-mass (green) and high-mass (pink) orbits that merge before z=0 shown as a function of redshift. 
    The horizontal line shows a merger fraction of 0.9, or 90\%. 
    High-mass pairs have merger fractions greater than 0.9 for $z>1.1$, and low-mass pairs for $z>1.6$.
    The sharp decline of the merger fraction to zero at low redshift is a non-physical feature of the simulation ending at $z=0$.
    Only orbits with sufficiently short merger timescales will merge at redshifts $z<\sim1$.}
    \label{fig:fmerge}
    \end{center}
\end{figure}

\subsection{Number of Pairs}\label{sec:pairprops-num}
The total number of pairs at a given redshift is equal to the number of orbits at that redshift, including both merger and non-merger pairs. 
A single, non-merging pair with first infall at $z=3$ will contribute to the number of pairs at all redshifts from $z=0-3$. 

Figure~\ref{fig:numorbits} shows the number of low-mass and high-mass pairs as a function of redshift from $z=0-6$. 
Low-mass pairs (green solid line) are most numerous between $z=1.25-2$, while high-mass pairs (pink solid line) are most numerous between $z=0-1$.
The dashed lines show the number of unique pairs that merge prior to $z=0$ for each sample. 
The number of pairs that merge decreases to zero at $z=0$ since many pairs that exist at low redshift will have merger timescales that span beyond the end of the simulation (i.e., $z=0$). 

As subhalos experience first infall into a FoF group, the number of pairs increases. 
On the other hand, mergers simultaneously lead to a decrease in the number of pairs.
The number of low-mass pairs increases from $\sim600$ pairs at $z=6$ to $\sim6,600$ at $z=2$. 
The decrease in the number of pairs after $z=2$ means that the number of low-mass pairs that are merging at each redshift is larger than the rate at which pairs are added to the sample.
The number of high-mass pairs also increases until $z=1$, at which point it remains approximately constant from $z=0-1$ at $\sim1000$ pairs. 
There is a slight decrease in the number of high mass pairs at the very lowest redshifts $z<0.1$, where the mergers begin to outnumber the new pairs being added at each redshift.

In Fig.~1 of \chambe{}, the number of low-mass pairs peaks (with $\sim3,000$ pairs) and begins to decrease at $z\sim2$, and the number of high-mass pairs is approximately constant between $z=0-1$, peaking at $\sim700$ pairs. 
We find the same behavior with our pair sample, with low-mass pairs peaking at z=2 and high mass pairs leveling off between $z=0-1$.  
However, in this study, the number of pairs at a given redshift is higher than in our previous work, since the orbit catalog includes a unique orbit for every pair from the previous work.
For example, a pair that passes the \chambe{} selection criteria only at $z=1$ will only be counted as a pair at $z=1$ in the previous work, while, in this study, it may be counted at many more redshifts since we can follow the orbit forwards and backwards in time. 

\subsection{Merger Fraction}\label{sec:pairprops-frac}
    We calculate the merger fraction by dividing the number of pairs that merge (dashed lines in Fig.~\ref{fig:numorbits}) by the total number of merging and non-merging orbits (solid lines) at a given redshift. 
    As before, an orbit with first infall at $z=2$ and merger at $z=1$ will be included in the merger fraction calculation for redshifts $z=1-2$.
    Note, this definition of the merger fraction differs from that typically used in observational studies, where merger fractions are computed from close pair fractions with a correction term~\cite[e.g.,][]{Ventou2019}. 
    
    Figure~\ref{fig:fmerge} shows the fraction of isolated pairs of low-mass (green) and high-mass (pink) pairs that merge before the end of the simulation as a function of redshift. 
    At redshifts $z>2$, the merger fraction for low-mass and high-mass pairs is greater than $0.9$.
    The merger fraction for both mass ranges declines to zero at $z=0$, due to the very small fraction of pairs at low redshift ($z<1$) that have short enough merger timescales to merge before $z=0$.
    
    The merger fraction of low-mass and high-mass pairs is the same from $z=0-0.5$ and $z=2.5-6$. 
    Between $z=0.5-2.5$, the high-mass merger fraction is larger than the low-mass merger fraction, and the knee of the merger fraction occurs at a lower redshift. 
    We discuss this difference in more detail in Sec.~\ref{sec:disc-nonmergers}.

    Note that the merger fraction defined here is a measure of the fraction of isolated pairs that merge before $z=0$. It is not, however, a measure of the fraction of \textit{all} low-mass and high-mass pairs in all environments that will merge.


\begin{figure*}[htb!]
    \begin{center}
    \includegraphics[width=0.7\textwidth]{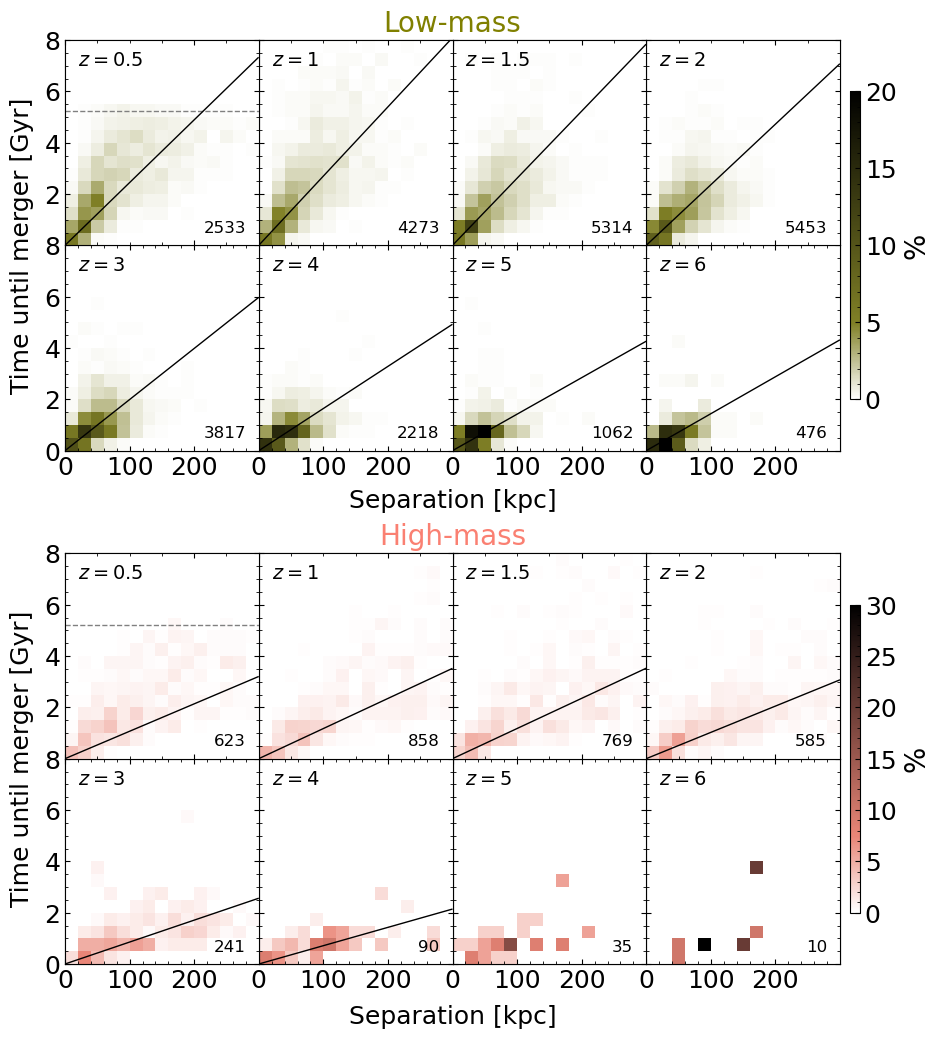}
    \caption{The distribution of merger timescales for low-mass (top) and high-mass (bottom) pairs as a function of physical separation at $z=0.5,1,1.5,2,3,4,5\mbox{and }6$. 
    The colorbars show the percentage of the total number of pairs at each redshift (given in the bottom right corner of each panel) that are in a given separation bin.
    In each of the first panels, the grey horizontal dashed line shows the time remaining in the simulation, above which there are no mergers. 
    The thin black line is the linear fit to each set of data with slopes from Tab.~\ref{tab:slopes} and described in more detail in the text.
    We find:
    i.) a positive correlation between separation and the time until merger at each selected redshift -- the lowest separation pairs tend to have the shortest merger timescales; 
    ii.) given a redshift and separation, the low-mass merger timescale is longer than the high-mass merger timescale, i.e. low-mass pairs at $z=1$ with separations $\rsep<100\,\kpc$ merge within $6\, \Gyr$, while high-mass pairs in the same separation range merge within $3\, \Gyr$; 
    iii.) merger timescales are longer for lower redshift pairs; 
    iv.) the spread of merger timescales and separations increase with decreasing redshift.}
    \label{fig:2dhist}
    \end{center}
\end{figure*}

\section{Results: The Mass and Redshift Dependence of Merger Timescales}\label{sec:results}
    Using our sample of isolated low-mass and high-mass pairs in the TNG100 simulation, we calculate the merger timescales, or time until merger, for all of the merging pairs in sample. 
    The merger timescale of a pair is defined to be the amount of time that elapses between the redshift at which pairs are selected and the merger redshift.
    Only the merging pairs will be considered for the remainder of the analysis. 

    In Sec.~\ref{sec:results-timevsep}, we explore how the merger timescale changes as a function of pair separation across redshifts for low-mass and high-mass pairs. 
    In Sec.~\ref{sec:results-timevredshift}, we investigate the median time until merger for all pairs as a function of redshift. 
    We will additionally examine the merger timescale's dependence on a variety of separation criteria. 
    Separation criteria are applied to pairs at each redshift independently, such that a pair in the $10-50\,\kpc$ bin at one redshift will not be part of the sample used at redshifts where its separation is $>50\,\kpc$.

\subsection{Separation Dependence of Merger Timescales}\label{sec:results-timevsep}
    We calculate the time until merger as a function of separation for low-mass and high-mass pairs at a variety of redshifts from $z=0.5-6$. 
    Binning the pairs by separation, we can study how the merger timescale changes for pairs selected at different points in their orbits. 
    
    We bin our pair sample by merger timescale, in bins of $0.5\Gyr$, and separation bins of $20\,\kpc$, and only consider the pairs with separations $\rsep>10\kpc$.
    Figure~\ref{fig:2dhist} shows a heatmap distribution of pairs as a function of merger timescale vs. separation for low-mass (top) and high-mass (bottom) pairs at redshifts $z=(0.5,1,1.5,2,3,4,5,6)$. 
    The colorbars to the right of each figure show the percentage of the population of pairs at that redshift that are in each bin.
    The number of pairs at each redshift is printed in the bottom right corner of each panel.
    The horizontal line in the first panel shows the time remaining in the simulation until $z=0$, above which no merging pairs exist.

    Additionally, each panel with more than 50 pairs includes a linear fit to the data prior to binning. 
    We fit a line of the form $y=C*x$, where $y$ is the time until merger, $x$ is separation, and $C$ is the slope that minimizes the mean squared error. 
    We do not include a fitting parameter for the intercept.
    The slope of each linear fit is given in Table~\ref{tab:slopes}.

    We find that the merger timescale for low-mass and high-mass pairs is positively correlated with the separation of the pair at a given redshift. 
    The slope of the linear fit increases with decreasing redshift, meaning that the merger timescale increases with decreasing redshift for a given separation.
    For example, pairs at $z=4$ with separations between $50-100\,\kpc$ tend to merge in less time than pairs with the same separations at $z=0.5$. 
    For the same physical separation, the merger timescale of high-mass pairs is shorter than for low-mass at each redshift. 
    
    In addition, the spreads in the merger timescales and pair separations are smaller at higher redshift, for both low- and high-mass pairs. 
    For example, the spread of merger timescales for low-mass pairs at $z=1$ (4273 pairs) goes from $0-8\,\Gyr$, while pairs at $z=4$ (2218 pairs) have merger timescales between $0-3\,\Gyr$. 
    Likewise, the spread of separations at $z=1$ is $0-200\kpc$, but at $z=4$ the spread is smaller, between $10-125\kpc$. 
    The high-mass pairs also have a larger spread in the distribution of merger timescales and separations at low redshift than at higher redshift.
    This is due to the growth of halos over time, and thus an increasing virial radius for FoF groups at lower redshifts. 

    Fig.~\ref{fig:2dhist} can be used to estimate the merger timescale of an isolated pair at a given redshift. 
    For example, a low-mass pair at $z=2$ with $\rsep\sim 75\,\kpc$ will merge in $0-5\,\Gyr$, with a most likely time to merger of around $2\Gyr$.
    On the other hand, a high-mass pair at $z=2$ with $\rsep\sim 75\,\kpc$ will merge in $0-2\,\Gyr$, with a most likely merger timescale of around $0-1\Gyr$.

    \input{table-slopes}

    %
    

\subsection{Redshift Dependence of Merger Timescales}\label{sec:results-timevredshift}
    In this subsection, we study the merger timescales of low-mass and high-mass pairs as a function of redshift. 
    
    First, we consider the pair sample as a whole, and quantify the median merger timescale for all merging low- and high-mass pairs from $z=0-6$. 
    We then consider two sets of separation criteria to create separation-selected subsamples, as in \chambe{}, and study the impact of different selection criteria on the resulting merger timescales.

    \subsubsection{Full sample}
\begin{figure}[tb]
            \centering
            \includegraphics[width=\columnwidth]{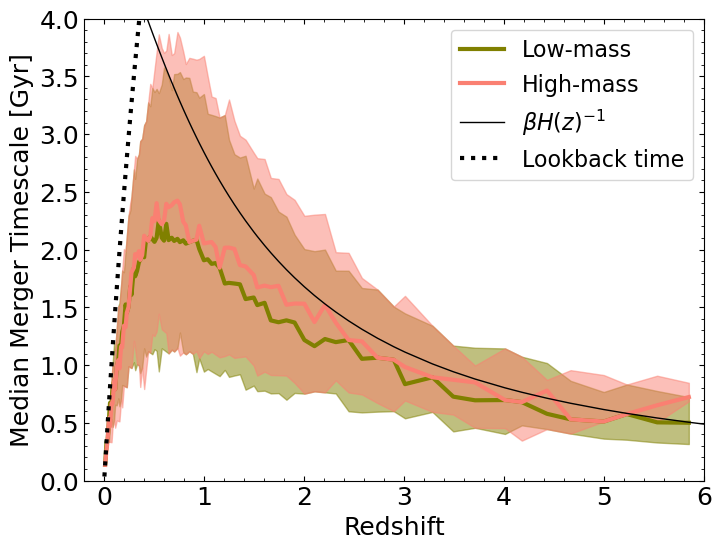}
            \caption{The median merger timescale as a function of redshift for low-mass (green) and high-mass (pink) pairs in our merger sample. 
            Shaded regions represent the 1st and 3rd quartile spread on the median. 
            The thick dotted black line at low redshifts shows the time to $z=0$ (the lookback time) as a function of redshift, which sets the upper bound for the merger timescale that a merging pair can have. 
            The thin black line shows the Hubble time as a function of redshift multiplied by a constant $\beta=0.35$, which approximates the median time until merger from $z=6$ to $z\sim1$.
            The median time until merger is similar for low-mass and high-mass pairs, and rises from $z=6$ to a peak at $z\sim0.75$, then decreases to $z=0$.}
            \label{fig:timescales}
        \end{figure}

        We calculate the merger timescales for pairs at each redshift, and quantify the median and spread on the merger timescale as a function of redshift in Fig.~\ref{fig:timescales}. As stated in  Section \ref{subsubsec:post-infall}, merger timescales begin at the time of first infall and conclude at the merger redshift (see Section \ref{subsec:mergers}).
        We include the full catalog of merging orbits at each redshift, only excluding separations $\rsep<10\kpc$ to limit the impact of subhalos becoming indistinguishable in the \subfind{} catalogs.

        Figure~\ref{fig:timescales} also shows
        the 1st and 3rd quartiles are indicated with shaded regions. 
        Low-mass merger timescales (green) and high-mass merger timescales (pink) are roughly equivalent at all redshifts, which implies that mergers do not proceed in fundamentally different ways at different mass scales. 
        We explore this point further in Sections~\ref{sec:results-phys} and~\ref{sec:results-scal}.

        The median time until merger is $\sim0.5-0.7\,\Gyr$ at $z\sim6$, then rises to a peak of $2.3-2.4\,\Gyr$ at $z\sim0.6$, then decreases to zero at $z=0$.
        The abrupt decrease is artificial rather than a true physical feature of merger timescales. 
        As $z\to0$, there is an increasingly small fraction of pairs that merge before $z=0$, as seen by the dashed lines in Fig~\ref{fig:numorbits}. 
        These mergers must proceed on shorter timescales by definition, as they were selected as pairs that must merge prior to $z=0$ (i.e., a pair at $z=0.1$ that merges by $z=0$ can have a maximal merger timescale of $1.47\Gyr$, whereas a pair at $z=1$ can have a maximal merger timescale of $8\Gyr$). 
        This is shown by the dotted black line on the left of Fig.~\ref{fig:timescales}, which shows the lookback time of the simulation as a function of redshift.\footnote{The lookback time at a given redshift $z_n$ is equivalent to the time elapsed between $z_n$ and $z=0$. Thus, merging pairs cannot have merger timescales larger than the lookback time of the simulation at a given redshift.} 

        To understand the functional form of merger timescales between $z=1-6$, which is shown by the thin black line in Fig.~\ref{fig:timescales}, we consider the following derivation. 
        In principle, merger timescales for a body moving through a homogeneous field of collision-less matter can be analytically derived from the Chandrasekhar formula for dynamical friction~\citep{Binney2008}.\footnote{Specifically, this derivation considers the merger timescale to be the time that elapses between the secondary subhalo crossing into the virial radius of the primary subhalo and the secondary's coalescence with the primary.} 
        Departures from the idealized case introduce perturbations to that solution, the validity of which has been tested in cosmological hydrodynamic and N-body simulations~\citep{Jiang2008, BoylanKolchin2008}. 

        Such studies have found that the merger timescale in N-body hydrodynamic simulations is of the form
        \begin{equation}
            \rm \tau_{\rm merge} = \frac{A(\Theta)}{\ln \Lambda}\frac{\mprim}{\msec}\tau_{\rm dyn}
        \end{equation}
        where $\mprim$ and $\msec$ are the primary and secondary subhalo masses, $A(\Theta)$ is a constant for a given orbital configuration, $\ln\Lambda$ is the Coulomb logarithm taken to be $\ln\Lambda = \ln(1+\mprim/\msec)$, and $\tau_{\rm dyn}$ is the dynamical timescale at the virial radius of the primary subhalo.
        The dynamical timescale is related to the crossing time at the virial radius, and is given by 
        \begin{equation}
            \rm \tau_{\rm dyn} = \frac{\Rvir}{V_{\rm circ}(\Rvir)},
        \end{equation}
        where $V_{\rm circ}(\Rvir)$ is the circular velocity at the virial radius of the primary subhalo. 
        Note that the dynamical time can thus be rewritten as 
        \begin{equation}
            \rm \tau_{\rm dyn} = (G\rho_{crit})^{-1/2} = \bigg(\frac{3\,H^2(z)}{8\pi}\bigg)^{-1/2}.
        \end{equation}

        In our study, we keep the stellar and FoF group mass criteria fixed as a function of redshift, such that for a given pair, the merger timescale scales with redshift as
        \begin{equation}
            \rm \tau_{merge} \propto \tau_{dyn} \propto H(z)^{-1},
        \end{equation}
        assuming that there is no (or weak) redshift dependence of the distribution of orbital parameters.
        Thus, the merger timescale is expected to scale with the Hubble time at a given redshift. 
        
        In Fig.~\ref{fig:timescales}, the black dashed line shows the Hubble time $1/H(z)$ multiplied by $\beta=0.35$. 
        We did not perform a fit for this multiplicative constant, since we are interested in investigating the behavior of the merger timescale as a function of redshift rather than the specific values of the merger timescale. 
        The redshift evolution of our median merger timescales is consistent with the scaling with $1/H(z)$ between $z=\sim1-6$.\footnote{$H(z)$ is calculated using the same cosmology as the TNG100 simulation, from~\citet{Planck2015}. Specifically, $\Omega_M=0.31$ and $\Omega_{\Lambda}=0.69$.}
        
       We note that the $1/H(z)$ functional form deviates from the median results at low redshift, but is still within the errors. The deviation follows the trend of decreasing merger fractions for the respective pair samples illustrated in Figure~\ref{fig:fmerge}. We expect that the deviation of the median results from $1/H(z)$ thus owes to insufficient time for all pairs to merge, rather than indicating that the true functional form should be shallower than $1/H(z)$.
        
        It is notable that such a simple scaling holds for a complex simulation like TNG100 where subhalos are far from isotropic spherical distributions of mass.
        It is particularly interesting as well that low-mass and high-mass pairs have the same merger timescales \textit{and} follow the $1/H(z)$ redshift scaling. 
        

    \subsubsection{Physical Separation Selected Pairs }
    \label{sec:results-phys}
        Since pair samples are typically picked via separation criteria in both simulations and observations, we study the impact of different separation-based selection criteria on inferred merger timescales.
        Separation selected samples have a lower separation criteria of $10\,\kpc$.
        This lower separation criteria is also commonly applied to observationally selected pairs in studies of merger fractions and merger rates~\citep[see e.g.][and observational studies cited therein]{Lotz2011,Snyder2017,Besla2018}.
        
        The first set of separation criteria selects only pairs at a given redshift that have physical 3D separations greater than $10\,\kpc$ and less than $[50, 70, 100, 150, 200, \mbox{and }300]\kpc$, yielding six pair subsamples. 
        Pairs that are selected via the $\rsep<50\kpc$ separation cut will be included in the $\rsep<70\kpc$ subsample, and so on.
        These separation criteria do not vary as a function of time or mass, and are applied equivalently to the low-mass and high-mass samples at all redshifts. 
        For example, an orbit in the separation bin $<50\,\kpc$ at $z=2$ will not necessarily be in that same bin at other redshifts. 
        
\begin{figure*}[htb]
        \centering
        \includegraphics[width=\textwidth]{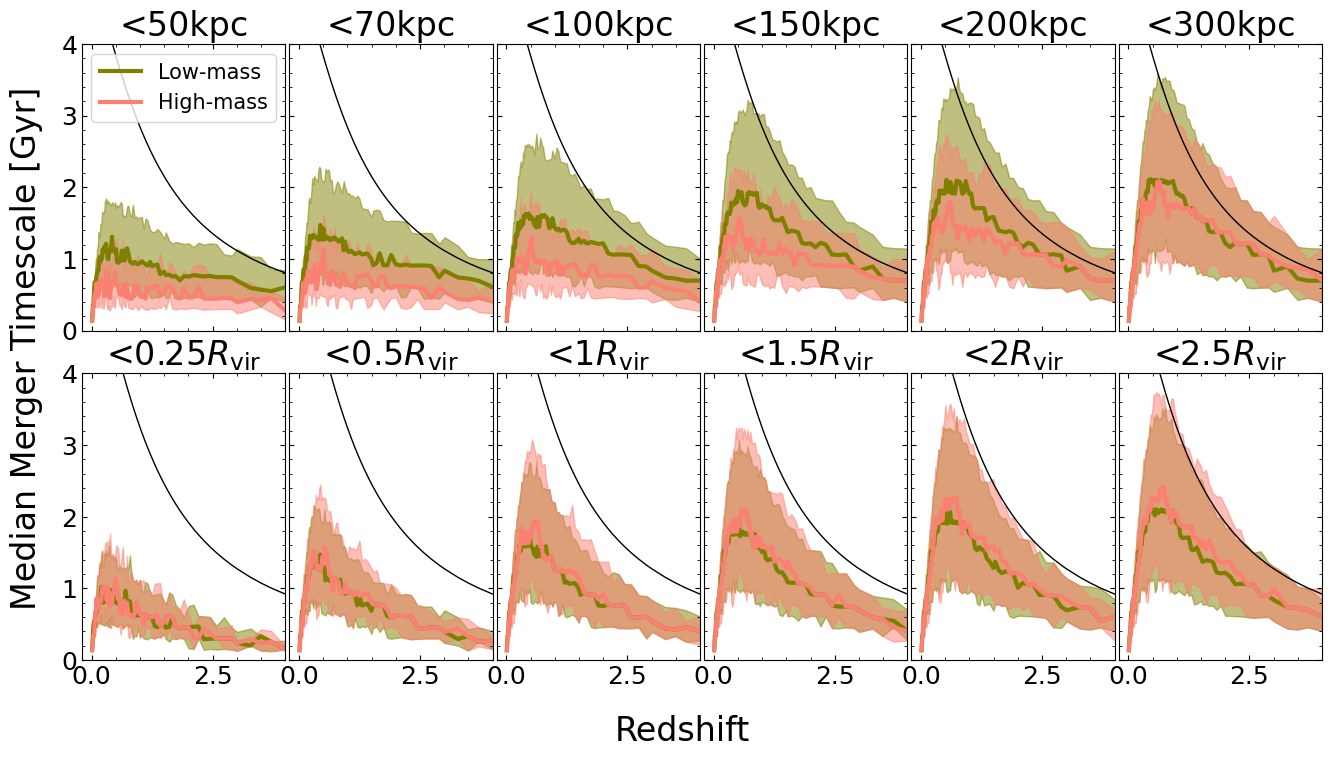}
        \caption{(Top) The median merger timescale as a function of redshift for low-mass (green) and high-mass (pink) pairs with 3D physical separations greater than $10\,\kpc$ and less than $50,70,100,150,200,\mbox{and }300\,\kpc$ from left to right. 
        (Bottom) The median time until merger as a function of redshift for pairs with physical separations greater than $10\,\kpc$ and less than $0.25, 0.5, 1, 1.5, 2,\mbox{and }2.5\,\Rvir$ from left to right. 
        The thin black line is the same as shown in Fig.~\ref{fig:timescales}, and goes as the Hubble time $1/H(z)$ multiplied by 0.35 (see Sec.~\ref{sec:results-timevredshift} for additional details).
        The time until merger increases from $z=4$ to $z\sim0.5-0.75$, at which point the time until merger decreases to zero since all low redshift mergers must have short merger timescales to merge before the end of the simulation at $z=0$. 
        }
        \label{fig:timescales-sep}
    \end{figure*}

        The top panel of Fig.~\ref{fig:timescales-sep} shows the merger timescale versus redshift for each of the low-mass (green) and high-mass (pink) pair subsamples with separations less than the physical separation listed above each panel. 
        The same Hubble time redshift scaling from Sec.~\ref{sec:results-timevredshift} is shown by the thin black line in each panel.
        We first note that the merger timescale peaks around redshift $z\sim0.5$ for all low-mass and high-mass subsamples. 
        In addition, all subsamples show the same decline of mean merger time as $z\to0$, as discussed in the previous section. 
        These traits are shown in Fig.~\ref{fig:timescales}, meaning they are features that are independent of separation criteria.
        
        
        The subsamples with maximum separations of $[50, 70, \mbox{and } 100]\,\kpc$ result in median merger timescales that are higher for low-mass pairs than high-mass pairs at all redshifts. 
        The difference in the merger timescale is up to $0.8\,\Gyr$ longer for low-mass pairs than high-mass pairs at the same redshift for the same separation cut. 
        The offset between the low-mass and high-mass merger timescale decreases for the largest separation cut of $\rsep<300\kpc$. 
        In the top rightmost panel, the median merger times converge for nearly all redshifts, similar to the merger timescales from the full sample shown in Fig. \ref{fig:timescales}.
        
        As the selection criteria increases (from left to right), each subsample contains a larger fraction of the full sample.
        The merger timescale for both low-mass and high-mass pairs increases with an increasing maximum separation cut.  
        This follows directly from Sec.~\ref{sec:results-timevsep}, where we found that the time until merger is positively correlated with increasing separation for all pairs. 
        Including a higher fraction of larger-separation systems in each subsample increases the median time until merger at all redshifts. 
        
        The median merger timescale for low-mass pairs does not change significantly for any separation cuts above $150\,\kpc$, which is larger than the average virial radius of low-mass systems at all redshifts. 
        On the other hand, high-mass merger timescales tend to increase as separation increases.
        High-mass pairs tend to have higher separations than low-mass pairs, as shown in Sec.~\ref{sec:results-timevsep}, due to the larger size of the high-mass subhalos. 
    \subsubsection{Scaled Separation Selected Pairs}
    \label{sec:results-scal}
        We calculate the merger timescales for six additional subsamples of pairs, applying separation criteria that scale with both redshift and mass.
        The scaled separation, which we define in Section~\ref{sec:methods}, is the separation of a pair divided by the virial radius of the pair's FoF group, $\Rvir$. 
        Scaled separation criteria select the equivalent \textit{fraction} of the full volume of each FoF group, regardless of mass or redshift, whereas physical separation criteria will select the same volume around each primary, with no account for the growth of dark matter halos or spatial distributions of satellites. 
        The scaled separation selected pairs have separations greater than $10\,\kpc$ and less than $[0.25, 0.5, 1, 1.5, 2, 2.5]\,\Rvir$. 
        As in the previous section, the selection criteria are applied at each redshift independently. 
        
        As detailed in \chambe{}, the median virial radius for low-mass FoF groups at $z=[0,1,2,3,4]$ is approximately $[134, 85, 59, 43, 33]\,\kpc$, and for high-mass FoF groups is approximately $[348, 206, 134, 97, 76]\,\kpc$.
        Thus, choosing a scaled separation cut of $\rsep<1\,\Rvir$ at $z=1$ will select low-mass pairs with separations $\rsep\lesssim85\,\kpc$ and high-mass pairs with separations $\rsep\lesssim206\,\kpc$.
        
        The bottom panel of Fig.~\ref{fig:timescales-sep} shows the median time until merger versus redshift for low-mass (green) and high-mass (pink) pair subsamples with scaled separations less than those listed at the top of each plot. 
        As in the top row, the thin black line in each panel shows the Hubble time scaling derived in Sec.~\ref{sec:results-timevredshift}.
        
        Quite unlike the physical separation selected subsamples in the top panels, using a scaled separation criteria results in nearly identical median merger timescales for low-mass and high-mass pairs at all redshifts.
        We find that the median time until merger at $z=1$ for each scaled separation cut is about $\sim[0.64, 0.94,1.25,1.55,1.85,1.96]\,\Gyr$  respectively, for both low-mass and high-mass samples.
        
        The median merger timescale starts off small at $z>4$, then increases to a peak around $z\sim0.6$, before quickly decreasing to zero at $z=0$. 
        This redshift evolution is similar to that of the full sample in Fig.~\ref{fig:timescales}.
        
        The median time until merger at a given redshift increases.
        The low-mass and high-mass merger timescales increase for separation criteria that include a larger fraction of the virial radius, and thus larger separation pairs, but recover the behavior of the full sample by a separation cut of $2\,\Rvir$.
        
        In addition, the slope of the merger timescale from $z=6\to1$ is steeper for high-mass pairs in the scaled separation subsamples than the physical separation subsamples.
        This is especially noticeable in the first panel of the top and bottom rows, where the merger timescale of high-mass pairs is approximately flat for the physical separation cut $\rsep<50\,\kpc$. 

\section{Discussion} \label{sec:discussion}
    In this section, we will explore the implications and broader impacts of our merger timescale analysis, and put our study in context with previous work. 

    In Sec.~\ref{sec:disc-pairfrac}, we return to one of the key questions posed in our previous work: what are the underlying physical mechanisms that drive the difference in pair fraction evolution between low- and high-mass pairs? 
    In Sec.~\ref{sec:disc-nonmergers}, we will discuss the non-merger portion of our orbit sample and the difference between the merger fractions in the low-mass and high-mass regimes.
    Finally, in Sec.~\ref{sec:disc-suggestions}, we discuss the implications of our work for future observational and theoretical merger fraction and merger rate studies and provide suggestions for a self-consistent framework moving forward.

    \subsection{Connections to Pair Fractions}\label{sec:disc-pairfrac}
        In our previous work~\citep{Chamberlain2024}, we found that the pair fractions of isolated low-mass and high-mass pairs evolve very differently from $z=0-2.5$ (see their Fig.~3). 
        Low-mass major pair fractions are constant for $z>2.2$, but decrease by almost 70\% from $z=3$ to $z=0$. 
        High-mass major pair fractions slowly increase from $z=6$ to $z=0$, with a more significant increase from $z=0.25$ to $z=0$.
        In short, the redshift evolution of pair fractions is opposite for low-mass pairs (which have a decreasing pair fraction with decreasing redshift) and high-mass pairs (which have an increasing pair fraction with decreasing redshift) from $z=3\to0$.
    
        \chambe{} suggested a hypothesis for the difference in the behavior of the pair fractions:  that perhaps low- mass and high-mass pairs have different merger timescales, which could lead to a difference in pair fractions from high to low redshift.
        
        However in the present study, we have shown that the merger timescales of the full sample of low-mass and high-mass pairs are approximately equal at all redshifts, meaning that the opposing behaviour of low-mass and high-mass pair fractions is not a result of different merger timescales. 
        Rather, we suspect that the build-up of larger structures due to hierarchical formation under $\Lambda$CDM leads to a reduced number of low-mass groups that enter the analysis at lower redshift. 
        Additionally through this study, we have found that a significant fraction of the pairs selected in our previous work do indeed merge ($>90\%$ of pairs for $z>1.6$), so that the rate at which mergers occur is likely larger than the rate at which isolated low-mass pairs form, particularly at low redshift, thus leading to the quickly declining pair fraction measured in our previous work.

    \subsection{Non-merger population}\label{sec:disc-nonmergers}
        In our catalog of 22,213 low-mass pairs and 3,029 high-mass pairs, only 3700 (16.66\%) low-mass pairs and 903 (29.71\%) high-mass pairs do not merge by $z=0$ (i.e. are classified here as ``non-mergers").
        In Sec.~\ref{sec:pairprops-frac}, we found that the merger fraction of low-mass and high-mass pairs at $z>3$ is $>0.95$ and the merger fraction decreases substantially at $z<3$. 
        While the decline in the merger fraction at $z\lesssim1$ is primarily due to the limited time that a pair will have to merge before the end of the simulation, we also find that between $z=\sim0.6-2.5$ the fraction of low-mass pairs is lower than that of high-mass pairs. 

        As in the previous subsection, one possible hypothesis that could explain the merger fraction differences is that low-mass pairs have shorter merger timescales than high-mass pairs, such that their mergers would occur at earlier times in the simulation leading to a decrease in the merger fraction at higher redshifts compared to high-mass pairs.
        However, we have shown that the merger timescales of our full low-mass and high-mass pair sample evolve similarly for $z=0-6$ which means that the difference in the low- and high-mass merger fraction between $z=0.5-2.5$ is not a result of a difference in merger timescales.
    
        We suspect that the cause of the difference between low- and high-mass merger fractions is not a difference in the merging population itself, but rather in the non-merger population and its evolution over time. 
        As we do not categorize our non-merging sample into ``likely mergers" and ``fly-by interactions" subsamples for this study (see Fig. \ref{fig:example-orbits}), we cannot comment on how these two populations individually contribute to the difference in merger fractions.
        However, one possible explanation could be that a larger fraction of infalling low-mass pairs at lower redshift results in fly-by interactions. 
        This could be the case if the velocity at which the secondary subhalo enters the primary's FoF group becomes increasingly large compared to the virial velocity\footnote{The virial velocity is the value of the circular velocity at the virial radius.}, resulting in fewer bound systems. 


    \subsection{Implications and Suggestions for Future Observational and Theoretical Studies of Pairs}~\label{sec:disc-suggestions}
        While we are unable to make direct comparisons with previous observational studies of merger fractions and merger rates, as our study uses true physical separations from the simulation rather than projected separations, we can still draw meaningful conclusions about the application of our results to future studies of merger timescales and merger rates. 
    
        \subsubsection{Pair Selection Criteria and Merger Rates in Illustris}
            Comparisons between studies of pair fractions, merger fractions, and merger rates, especially across cosmic time and different mass scales, are often challenged by the implementation of different selection criteria in each study.
            
            In some observational studies, pair selection criteria are often set by the observational parameters of the survey itself. 
            For example, limiting magnitudes and completeness limits can dictate which range of stellar masses and stellar mass ratios are considered~\citep{Patton2000,Lotz2008,Man2016,Ventou2019}. Additionally, specific separation criteria may be chosen to avoid fiber collisions in a spectroscopic survey~\citep[as in][]{Patton2011,Besla2018}.
        
            In theoretical studies, it is more straightforward to adopt several different pair selection criteria simultaneously for comparisons to specific observational studies~\citep[i.e.][]{Lotz2011,RG2015,Besla2018,Snyder2023}. 
            However, this then limits theoretical studies to a subset of pair selection criteria specific to their target comparison studies.
    
            In recent years, some work has aimed to standardize pair selection criteria for future studies, in particular, to facilitate a fair comparison between theory and observations. 
            One such study is~\citet{Ventou2019}, which created mock catalogs from the original Illustris-1 simulation to analyze the merger probability of galaxy pairs as a function of (both physical and projected) separation and relative velocity. 
            They find that selecting pairs with projected separations between $5\leq r_{\rm proj} \leq 50\,\kpc$ and projected relative velocities of $v_{\rm proj}\leq300\,\kms$ selects all pairs that have greater than 30\% chance of merging before $z=0$. 
            These selection criteria are calibrated to select a population of pairs that are fairly likely to merge.
            Using their projected separation vs velocity analysis of merger probabilities, they also develop a weighting scheme to determine the likelihood of merger as a function of projected separation and velocity, which provides a way to more accurately calibrate merger fractions from close pair fractions. 
            While their study considers the impact of stellar mass and redshift on the probability of merger, their suggested pair selection criteria do not vary with mass or redshift, which we have shown is crucial for interpreting and comparing pair fractions and merger timescales. 

                       
            \citet{RG2015} constructed galaxy merger trees to calculate the galaxy merger rate as a function of the stellar mass of the descendent subhalo and redshift in the original Illustris simulation. 
            They find that the galaxy merger rate increases with increasing redshift for descendent galaxy masses $\ms{}>10^{10}\,\Msun$. 
            These findings are consistent with predictions of the galaxy merger rate from semi-empirical models~\citep{Stewart2009,Hopkins2010a}, and with merger rates derived from observational merger fraction measurements calibrated by the corrected observability timescales of \citet{Lotz2011}.
            
            Additionally, \citet{RG2015} find that the major merger rate at $z\sim0.1$ increases with increasing descendant subhalo mass.
            While the redshift evolution of merger rates are broadly consistent with observations, they find significant discrepancies with major merger rates of galaxies with $\ms\, <10^{10}\,\Msun$ from~\citet{Casteels2014}, which used morphological merger signatures to estimate merger rates at low $z$. 
            As the~\citet{RG2015} work is self-consistently carried out at all redshifts and descendant masses, \citet{RG2015} suggest that the discrepancy arises from significant uncertainty in the observability timescales of galaxies with lower mass $\ms{}<10^{10}\,\Msun$, since the timescales in \citet{Casteels2014} are calibrated using an extrapolation of gas fraction data that is only available for $\ms{}>10^{10}\,\Msun$. 
            
            These results imply that the timescales used to convert observational merger fractions (determined either by weighted close pair fractions or by morphologically selected pairs) are sensitive to the mass ranges and redshifts at which they are valid, and attempting to make comparisons between theoretical predictions and observational measurements can lead to seemingly incompatible results. 
            Our work provides a unique framework that will allow for more robust comparisons of theoretical and observational samples. 

        \subsubsection{Implementing Scaled Separation Pair Selection Criteria in Future Studies}
            In our previous work, we studied the impact of different separation criteria on the recovered pair fractions of low-mass and high-mass pairs, and found that features of the redshift evolution of the pair fractions can only be distinguished by employing scaled separation criteria~\citep{Chamberlain2024}. 
            In the present work, we explore how these same set of selection criteria impact the recovered merger timescales. 
            We find that the median merger timescales of all isolated low-mass and high-mass pairs evolve nearly identically between $z=0-6$.
            All subsamples of our pair catalog from scaled separation cuts result in identical behavior of the low-mass and high-mass merger timescales as a function of redshift as well. 
            On the other hand, selecting pairs using a static physical separation criteria results in merger timescales that evolve differently between the two mass scales, and can differ by $0.8\Gyr$. 
            
            
            These results translate to important implications for observational studies of merger rates, which are typically determined using physical separation-selected close pairs and merger timescales. 
            Our work has shown that both of these quantities can vary significantly for different pair selection criteria and that physical separation criteria can eliminate the distinguishing features of pair fractions or the equivalence of merger timescales between low-mass and high-mass pairs.
             
            We therefore promote the adoption of separation criteria that vary as a function of mass and redshift for future pair studies. 
            In particular, we point out the importance of utilizing pair selection criteria that permit fair comparisons of pair properties in observational studies (i.e. pair fractions, merger timescales, merger rates, etc.), particularly when the goal of such a study is to quantify and compare the redshift evolution of pair properties at different mass scales.
            
            Specifically, we suggest employing separation criteria based on the scaled separation of each pair, given in Equation~\ref{eq:scaled}. 
            The application of a scaled separation as demonstrated here can be applied to observational studies. 
            Since estimates of a pair's stellar masses are needed to quantify the mass ratio of the pair (which is commonly computed for merger rate studies), no further observational information is needed to develop mass and redshift evolving separation cuts. 
            
            We provide a step by step example of this application as follows. First, compute the associated dark matter masses from the observed stellar masses of the pair using a SMHM relation~\citep[in this study, we employ that of][]{Moster2013}.
            Next, calculate the virial radius of a dark matter halo with a virial mass equal to the sum of the dark matter masses of the pair.\footnote{We showed in~\chambe{} that the virial radius of a halo with mass given by the combined dark matter halo masses of the primary and secondary recovers the virial radius of the FoF group with approximately 98\% accuracy. The virial radius can be calculated from
            \begin{equation}\label{eq}
                \Rvir = \sqrt[3]{\frac{3\rm M_{vir}}{4\pi \Delta_{c}\rho_{c}}},
            \end{equation} 
            where $\rho_c$ is the critical density of the Universe, and $\Delta_{c}$ is the overdensity constant~\citep[see][]{Binney2008}.}
            Finally, determine the physical separation for each pair individually that corresponds to the scaled separation criteria adopted in the study.

            The process of computing the dark matter halo masses of the observed galaxies typically introduces systematic error to an observational study.
            When converting from stellar mass to dark matter halo mass, the SMHM relationship can be sampled, i.e. by computing many realizations of the dark matter mass of each galaxy, in a similar fashion as performed in~\chambe{}, to derive the associated spread introduced by the abundance matching process.
            
            As a concluding note, we remind readers that our results are explicitly for isolated pairs, as outlined in Sec.~\ref{sec:methods}. 
            One strength of this approach is that we have been able to study dynamics that are inherent to the pairs themselves, rather than those that are a product of the environment. 
            Low-mass and high-mass pair merger timescales in high density environments may evolve differently than in our findings. 
            We leave the extension of this analysis to more high-density environments as the focus of future work. 
   
\section{Summary and Conclusions}\label{sec:conclusions}
    In this study, we construct a sample of the orbits of isolated low-mass ($10^8 <\ms{}<5\times10^{9}\Msun$) and high-mass ($5\times10^{9}<\ms{}<10^{11}\Msun$) major pairs (stellar mass ratio $> 1:4$) from $z=0-6$ in the TNG100 simulation.
    Orbits of pairs, i.e. the 3D physical separation between the pair as a function of time, are defined from the redshift at which the primary and secondary subhalo first share a common FoF group (i.e., `first infall') to either $z=0$ or the redshift at which the pair merges.
    
    The sample consists of 22,213 unique low-mass orbits and 3,029 unique-high mass orbits, for which we quantify the merger fraction as a function of redshift.
    We calculate the merger timescales of low-mass and high-mass major pairs as a function of separation and separately as a function of redshift. 
    Our goal is to identify the merger timescales of pairs in a cosmological framework and to compare the redshift evolution of the merger timescale between low-mass and high-mass pairs. 
    
    Additionally, motivated by our previous work on the pair fractions of low-mass and high-mass pairs in \citet{Chamberlain2024} where we showed that the evolution of the pair fraction is sensitive to the separation criteria used to select the pairs, we seek to determine the corresponding impact of various selection criteria on the merger timescales of pairs. 
    Specifically, we look at two sets of selection criteria, one which selects pairs via a static physical separation cut, and the other which selects pairs based on a separation cut that evolves with redshift and mass.
    This is especially important for studies that seek to study pair fractions, merger fractions, merger timescales, and merger rates at different mass scales and/or as a function of redshift. 

    Our main conclusions are as follows: 
    \begin{itemize}
        \item The merger fraction of physically associated low-mass and high-mass pairs is high ($>0.9$) at $z\gtrsim1.5$ (see Fig.~\ref{fig:fmerge}). However, the merger fraction declines rapidly to zero as $z\to0$, due to the finite length of the simulation, which artificially reduces the number of pairs with enough time left to merge at low redshifts. 
        \item The merger timescale of a pair at a given redshift increases with increasing pair separation. Additionally, for a given physical separation, high-mass pairs have a shorter merger timescale than low-mass pairs at the same redshift. For example, low-mass pairs at $z=1$ with separations $\rsep<100\,\kpc$ merge within $6\, \Gyr$, while high-mass pairs in the same separation range merge within $3\, \Gyr$ (see Fig.~\ref{fig:2dhist}).
        \item 
        The median merger timescale peaks around $z\sim0.6$ at $\sim2.1-2.4\Gyr$ for both low-mass and high-mass pairs.
        At redshifts $z\gtrsim1$, the median merger timescale for the full sample of low-mass and high-mass pairs that merge prior to $z=0$ declines with increasing redshift at a rate proportional to $1/H(z)$ (see Fig.~\ref{fig:timescales}). 
        At $z\lesssim1$, the maximum merger timescale is constrained by the time remaining in the simulation for the pair to merge, which causes the median merger timescale to decline to zero at $z=0$. 
        The decrease in the merger time at low redshifts is artificial and not representative of the true merger timescales of pairs at low redshift.
        \item When pairs are selected via a scaled separation criteria, namely the pair separation scaled by the virial radius $\Rvir$ of the FoF group of the primary, low-mass and high-mass merger timescales are nearly identical at all redshifts. This holds for all scaled separation criteria considered ($0.25,0.5,1,1.5,2,\mbox{and }3\,\Rvir$). At $z=[0, 1, 2, 3, 4]$, a scaled separation of $1\,\Rvir$ corresponds to an average physical separation of $[134, 85, 59, 43, 33]\kpc$ for low-mass pairs and $[348, 206, 134, 97, 76] \kpc$ for high-mass pairs.
        \item When pairs are selected via a physical separation criteria (one that does not vary with redshift or with mass), the median merger timescales of low-mass and high-mass pairs differ by up to $0.8\Gyr$. Thus, using the same merger timescale for low-mass and high-mass close pair samples selected via the same physical separation cuts will result in biased merger rate estimates.
    \end{itemize}

    Studies have found that merger rates of galaxy pairs vary with redshift and with the mass of the pair~\citep{Stewart2009,Hopkins2010a,RG2015}.
    In our previous work, we found that the pair fraction of isolated pairs in TNG100 likewise evolves with redshift and mass~\citep{Chamberlain2024}. 
    In this paper, we show that the merger timescales of major pairs in TNG100 vary with redshift, but that low-mass and high-mass pairs have equal merger timescales at all redshifts from $z=0-6$ if the correct separation selection criteria is used to pick equivalent samples of pairs.
    
    These works together give a comprehensive framework for inferring merger rates via pair fractions and merger timescales at redshifts from at least $z=0-6$, and for quantifying the differences between low-mass and high-mass pairs.
    Indeed, separation selection criteria that scale with the mass and redshift of the target system are crucial for interpreting pair properties in a self-consistent way. 
    In the future, observatories such as the Rubin Observatory, the Roman Space Telescope, JWST, and future ELTs, will detect an abundance of low-mass pairs at a wide range of redshifts, and our theoretical framework for interpreting these observations will be more critical than ever. 

\section*{Acknowledgements}
K.C. and G.B. are supported by NSF CAREER award AST-1941096. 
E.P. acknowledges financial support provided by NASA through the NASA Hubble Fellowship grant \#HST-HF2-51540.001-A awarded by the Space Telescope Science Institute, which is operated by the Association of Universities for Research in Astronomy, Incorporated, under NASA contract NAS5-26555.
P.T. acknowledges support from NSF-AST 2346977.

The \textsc{IllustrisTNG} simulations were undertaken with compute time awarded by the Gauss Centre for Supercomputing (GCS) under GCS Large-Scale Projects GCS-ILLU and GCS-DWAR on the GCS share of the supercomputer Hazel Hen at the High Performance Computing Center Stuttgart (HLRS), as well as on the machines of the Max Planck Computing and Data Facility (MPCDF) in Garching, Germany.

This work is based upon High Performance Computing (HPC) resources supported by the University of Arizona TRIF, UITS, and Research, Innovation, and Impact (RII) and maintained by the UArizona Research Technologies department.

We respectfully acknowledge the University of Arizona is on the land and territories of Indigenous peoples. Today, Arizona is home to 22 federally recognized tribes, with Tucson being home to the O’odham and the Yaqui. Committed to diversity and inclusion, the University strives to build sustainable relationships with sovereign Native Nations and Indigenous communities through education offerings, partnerships, and community service.

\bibliography{refs}{}
\bibliographystyle{aasjournal}

\end{document}

%% file: math_definitions.tex
\newcommand{\Msun}{\ensuremath{\mathrm{M}_\odot}}

\newcommand{\kms}{\ensuremath{\mathrm{km}~\mathrm{s}^{-1}}}

\newcommand{\kpc}{\ensuremath{\mathrm{\,kpc}}}
\newcommand{\Mpc}{\ensuremath{\mathrm{Mpc}}}

\newcommand{\Myr}{\ensuremath{\mathrm{Myr}}}
\newcommand{\Gyr}{\ensuremath{\mathrm{\,Gyr}}}

\newcommand{\mprim}{\ensuremath{\rm M_{prim}}}
\newcommand{\msec}{\ensuremath{\rm M_{sec}}}

\newcommand{\paircat}{\textit{Full Pair Catalog}}

\newcommand{\lcdm}{\ensuremath{\Lambda \rm CDM}} 
\newcommand{\sublink}{\textsc{sublink}} 
\newcommand{\subfind}{\textsc{subfind}} 

\newcommand{\Mpeak}{\ensuremath{M_{\mathrm{peak}}}}
\newcommand{\Mhalo}{\ensuremath{M_{\mathrm{h}}}}
\newcommand{\MG}{\ensuremath{M_{\mathrm{G}}}}

\newcommand{\ms}[1]{\ensuremath{M_{*{#1}}}}

\newcommand{\Rvir}{\ensuremath{R_{\mathrm{vir}}}}

\newcommand{\rsep}{\ensuremath{r_{\mathrm{sep}}}}

%% file: table-slopes.tex
\begin{table}[tbp]
\centering
\caption{The slope of the linear fit between the time until merger as a function of separation for low-mass and high-mass pairs at a collection of redshifts. The corresponding line-of-best-fit is shown in Fig.~\ref{fig:2dhist}.}
\begin{tabular}{c|c|c}
\hline \hline
Redshift & Low-mass slope & High-mass slope\\\hline
0.5 & 0.024 & 0.011 \\
1.0 & 0.027 & 0.012 \\
1.5 & 0.026 & 0.012 \\
2.0 & 0.024 & 0.010 \\
3.0 & 0.020 & 0.009 \\
4.0 & 0.016 & 0.007 \\
5.0 & 0.014 & -- \\
6.0 & 0.014 & -- \\
 \hline \hline
\end{tabular}
\label{tab:slopes}
\end{table}

%% file: main.bbl
\begin{thebibliography}{}
\expandafter\ifx\csname natexlab\endcsname\relax\def\natexlab#1{#1}\fi
\providecommand{\url}[1]{\href{#1}{#1}}
\providecommand{\dodoi}[1]{doi:~\href{http://doi.org/#1}{\nolinkurl{#1}}}
\providecommand{\doeprint}[1]{\href{http://ascl.net/#1}{\nolinkurl{http://ascl.net/#1}}}
\providecommand{\doarXiv}[1]{\href{https://arxiv.org/abs/#1}{\nolinkurl{https://arxiv.org/abs/#1}}}

\bibitem[{{Behroozi} {et~al.}(2020){Behroozi}, {Conroy}, {Wechsler}, {Hearin},
  {Williams}, {Moster}, {Yung}, {Somerville}, {Gottl{\"o}ber}, {Yepes}, \&
  {Endsley}}]{Behroozi2020}
{Behroozi}, P., {Conroy}, C., {Wechsler}, R.~H., {et~al.} 2020, \mnras, 499,
  5702, \dodoi{10.1093/mnras/staa3164}

\bibitem[{{Besla} {et~al.}(2018){Besla}, {Patton}, {Stierwalt},
  {Rodriguez-Gomez}, {Patel}, {Kallivayalil}, {Johnson}, {Pearson}, {Privon},
  \& {Putman}}]{Besla2018}
{Besla}, G., {Patton}, D.~R., {Stierwalt}, S., {et~al.} 2018, \mnras, 480,
  3376, \dodoi{10.1093/mnras/sty2041}

\bibitem[{{Binney} \& {Tremaine}(2008)}]{Binney2008}
{Binney}, J., \& {Tremaine}, S. 2008, {Galactic Dynamics: Second Edition}
  ({Princeton University Press})

\bibitem[{{Boylan-Kolchin} {et~al.}(2008){Boylan-Kolchin}, {Ma}, \&
  {Quataert}}]{BoylanKolchin2008}
{Boylan-Kolchin}, M., {Ma}, C.-P., \& {Quataert}, E. 2008, \mnras, 383, 93,
  \dodoi{10.1111/j.1365-2966.2007.12530.x}

\bibitem[{{Bryan} \& {Norman}(1998)}]{Bryan1998}
{Bryan}, G.~L., \& {Norman}, M.~L. 1998, \apj, 495, 80, \dodoi{10.1086/305262}

\bibitem[{{Byrne-Mamahit} {et~al.}(2024){Byrne-Mamahit}, {Patton}, {Ellison},
  {Bickley}, {Ferreira}, {Hani}, {Quai}, \& {Wilkinson}}]{ByrneMamahit2024}
{Byrne-Mamahit}, S., {Patton}, D.~R., {Ellison}, S.~L., {et~al.} 2024, \mnras,
  528, 5864, \dodoi{10.1093/mnras/stae419}

\bibitem[{{Casteels} {et~al.}(2014){Casteels}, {Conselice}, {Bamford},
  {Salvador-Sol{\'e}}, {Norberg}, {Agius}, {Baldry}, {Brough}, {Brown},
  {Drinkwater}, {Driver}, {Graham}, {Bland-Hawthorn}, {Hopkins}, {Kelvin},
  {L{\'o}pez-S{\'a}nchez}, {Loveday}, {Robotham}, \&
  {V{\'a}zquez-Mata}}]{Casteels2014}
{Casteels}, K. R.~V., {Conselice}, C.~J., {Bamford}, S.~P., {et~al.} 2014,
  \mnras, 445, 1157, \dodoi{10.1093/mnras/stu1799}

\bibitem[{{Chamberlain} {et~al.}(2023){Chamberlain}, {Price-Whelan}, {Besla},
  {Cunningham}, {Garavito-Camargo}, {Pe{\~n}arrubia}, \&
  {Petersen}}]{Chamberlain2023}
{Chamberlain}, K., {Price-Whelan}, A.~M., {Besla}, G., {et~al.} 2023, \apj,
  942, 18, \dodoi{10.3847/1538-4357/aca01f}

\bibitem[{{Chamberlain} {et~al.}(2024){Chamberlain}, {Besla}, {Patel},
  {Rodriguez-Gomez}, {Torrey}, {Martin}, {Johnson}, {Kallivayalil}, {Patton},
  {Pearson}, {Privon}, \& {Stierwalt}}]{Chamberlain2024}
{Chamberlain}, K., {Besla}, G., {Patel}, E., {et~al.} 2024, \apj, 962, 162,
  \dodoi{10.3847/1538-4357/ad19d0}

\bibitem[{{Dolag} {et~al.}(2009){Dolag}, {Borgani}, {Murante}, \&
  {Springel}}]{Dolag2009}
{Dolag}, K., {Borgani}, S., {Murante}, G., \& {Springel}, V. 2009, \mnras, 399,
  497, \dodoi{10.1111/j.1365-2966.2009.15034.x}

\bibitem[{{Garavito-Camargo} {et~al.}(2019){Garavito-Camargo}, {Besla},
  {Laporte}, {Johnston}, {G{\'o}mez}, \& {Watkins}}]{GaravitoCamargo2019}
{Garavito-Camargo}, N., {Besla}, G., {Laporte}, C. F.~P., {et~al.} 2019, \apj,
  884, 51, \dodoi{10.3847/1538-4357/ab32eb}

\bibitem[{{Gardner} {et~al.}(2006){Gardner}, {Mather}, {Clampin}, {Doyon},
  {Greenhouse}, {Hammel}, {Hutchings}, {Jakobsen}, {Lilly}, {Long}, {Lunine},
  {McCaughrean}, {Mountain}, {Nella}, {Rieke}, {Rieke}, {Rix}, {Smith},
  {Sonneborn}, {Stiavelli}, {Stockman}, {Windhorst}, \& {Wright}}]{Gardner2006}
{Gardner}, J.~P., {Mather}, J.~C., {Clampin}, M., {et~al.} 2006, \ssr, 123,
  485, \dodoi{10.1007/s11214-006-8315-7}

\bibitem[{{G{\'o}mez} {et~al.}(2015){G{\'o}mez}, {Besla}, {Carpintero},
  {Villalobos}, {O'Shea}, \& {Bell}}]{Gomez2015}
{G{\'o}mez}, F.~A., {Besla}, G., {Carpintero}, D.~D., {et~al.} 2015, \apj, 802,
  128, \dodoi{10.1088/0004-637x/802/2/128}

\bibitem[{{Guzm{\'a}n-Ortega} {et~al.}(2023){Guzm{\'a}n-Ortega},
  {Rodriguez-Gomez}, {Snyder}, {Chamberlain}, \&
  {Hernquist}}]{GuzmanOrtega2023}
{Guzm{\'a}n-Ortega}, A., {Rodriguez-Gomez}, V., {Snyder}, G.~F., {Chamberlain},
  K., \& {Hernquist}, L. 2023, \mnras, 519, 4920,
  \dodoi{10.1093/mnras/stac3334}

\bibitem[{{Hopkins} {et~al.}(2010{\natexlab{a}}){Hopkins}, {Croton}, {Bundy},
  {Khochfar}, {van den Bosch}, {Somerville}, {Wetzel}, {Keres}, {Hernquist},
  {Stewart}, {Younger}, {Genel}, \& {Ma}}]{Hopkins2010b}
{Hopkins}, P.~F., {Croton}, D., {Bundy}, K., {et~al.} 2010{\natexlab{a}}, \apj,
  724, 915, \dodoi{10.1088/0004-637x/724/2/915}

\bibitem[{{Hopkins} {et~al.}(2010{\natexlab{b}}){Hopkins}, {Bundy}, {Croton},
  {Hernquist}, {Keres}, {Khochfar}, {Stewart}, {Wetzel}, \&
  {Younger}}]{Hopkins2010a}
{Hopkins}, P.~F., {Bundy}, K., {Croton}, D., {et~al.} 2010{\natexlab{b}}, \apj,
  715, 202, \dodoi{10.1088/0004-637X/715/1/202}

\bibitem[{{Jiang} {et~al.}(2008){Jiang}, {Jing}, {Faltenbacher}, {Lin}, \&
  {Li}}]{Jiang2008}
{Jiang}, C.~Y., {Jing}, Y.~P., {Faltenbacher}, A., {Lin}, W.~P., \& {Li}, C.
  2008, \apj, 675, 1095, \dodoi{10.1086/526412}

\bibitem[{{Kado-Fong} {et~al.}(2024){Kado-Fong}, {Robinson}, {Nyland},
  {Greene}, {Suess}, {Stierwalt}, \& {Beaton}}]{KadoFong2024}
{Kado-Fong}, E., {Robinson}, A., {Nyland}, K., {et~al.} 2024, \apj, 963, 37,
  \dodoi{10.3847/1538-4357/ad18cb}

\bibitem[{{Kristensen} {et~al.}(2021){Kristensen}, {Pimbblet}, {Gibson},
  {Penny}, \& {Koudmani}}]{Kristensen2021}
{Kristensen}, M.~T., {Pimbblet}, K.~A., {Gibson}, B.~K., {Penny}, S.~J., \&
  {Koudmani}, S. 2021, \apj, 922, 127, \dodoi{10.3847/1538-4357/ac236d}

\bibitem[{{Lotz} {et~al.}(2011){Lotz}, {Jonsson}, {Cox}, {Croton}, {Primack},
  {Somerville}, \& {Stewart}}]{Lotz2011}
{Lotz}, J.~M., {Jonsson}, P., {Cox}, T.~J., {et~al.} 2011, \apj, 742, 103,
  \dodoi{10.1088/0004-637x/742/2/103}

\bibitem[{{Lotz} {et~al.}(2008){Lotz}, {Davis}, {Faber}, {Guhathakurta},
  {Gwyn}, {Huang}, {Koo}, {Le Floc'h}, {Lin}, {Newman}, {Noeske}, {Papovich},
  {Willmer}, {Coil}, {Conselice}, {Cooper}, {Hopkins}, {Metevier}, {Primack},
  {Rieke}, \& {Weiner}}]{Lotz2008}
{Lotz}, J.~M., {Davis}, M., {Faber}, S.~M., {et~al.} 2008, \apj, 672, 177,
  \dodoi{10.1086/523659}

\bibitem[{{Luber} {et~al.}(2022){Luber}, {Pearson}, {Putman}, {Besla},
  {Stierwalt}, \& {Meyers}}]{Luber2022}
{Luber}, N., {Pearson}, S., {Putman}, M.~E., {et~al.} 2022, \aj, 163, 49,
  \dodoi{10.3847/1538-3881/ac3750}

\bibitem[{{Man} {et~al.}(2016){Man}, {Zirm}, \& {Toft}}]{Man2016}
{Man}, A. W.~S., {Zirm}, A.~W., \& {Toft}, S. 2016, \apj, 830, 89,
  \dodoi{10.3847/0004-637x/830/2/89}

\bibitem[{{Marinacci} {et~al.}(2018){Marinacci}, {Vogelsberger}, {Pakmor},
  {Torrey}, {Springel}, {Hernquist}, {Nelson}, {Weinberger}, {Pillepich},
  {Naiman}, \& {Genel}}]{TNG2}
{Marinacci}, F., {Vogelsberger}, M., {Pakmor}, R., {et~al.} 2018, \mnras, 480,
  5113, \dodoi{10.1093/mnras/sty2206}

\bibitem[{{Martin} {et~al.}(2018){Martin}, {Kaviraj}, {Devriendt}, {Dubois}, \&
  {Pichon}}]{Martin2018}
{Martin}, G., {Kaviraj}, S., {Devriendt}, J.~E.~G., {Dubois}, Y., \& {Pichon},
  C. 2018, \mnras, 480, 2266, \dodoi{10.1093/mnras/sty1936}

\bibitem[{{Martin} {et~al.}(2021){Martin}, {Jackson}, {Kaviraj}, {Choi},
  {Devriendt}, {Dubois}, {Kimm}, {Kraljic}, {Peirani}, {Pichon}, {Volonteri},
  \& {Yi}}]{Martin2021}
{Martin}, G., {Jackson}, R.~A., {Kaviraj}, S., {et~al.} 2021, \mnras, 500,
  4937, \dodoi{10.1093/mnras/staa3443}

\bibitem[{{Martin} {et~al.}(2022){Martin}, {Bazkiaei}, {Spavone}, {Iodice},
  {Mihos}, {Montes}, {Benavides}, {Brough}, {Carlin}, {Collins}, {Duc},
  {G{\'o}mez}, {Galaz}, {Hern{\'a}ndez-Toledo}, {Jackson}, {Kaviraj}, {Knapen},
  {Mart{\'\i}nez-Lombilla}, {McGee}, {O'Ryan}, {Prole}, {Rich}, {Rom{\'a}n},
  {Shah}, {Starkenburg}, {Watkins}, {Zaritsky}, {Pichon}, {Armus}, {Bianconi},
  {Buitrago}, {Bus{\'a}}, {Davis}, {Demarco}, {Desmons}, {Garc{\'\i}a},
  {Graham}, {Holwerda}, {Hon}, {Khalid}, {Klehammer}, {Klutse}, {Lazar},
  {Nair}, {Noakes-Kettel}, {Rutkowski}, {Saha}, {Sahu}, {Sola},
  {V{\'a}zquez-Mata}, {Vera-Casanova}, \& {Yoon}}]{Martin2022}
{Martin}, G., {Bazkiaei}, A.~E., {Spavone}, M., {et~al.} 2022, \mnras, 513,
  1459, \dodoi{10.1093/mnras/stac1003}

\bibitem[{{Moster} {et~al.}(2013){Moster}, {Naab}, \& {White}}]{Moster2013}
{Moster}, B.~P., {Naab}, T., \& {White}, S.~D.~M. 2013, \mnras, 428, 3121,
  \dodoi{10.1093/mnras/sts261}

\bibitem[{{Naiman} {et~al.}(2018){Naiman}, {Pillepich}, {Springel},
  {Ramirez-Ruiz}, {Torrey}, {Vogelsberger}, {Pakmor}, {Nelson}, {Marinacci},
  {Hernquist}, {Weinberger}, \& {Genel}}]{TNG4}
{Naiman}, J.~P., {Pillepich}, A., {Springel}, V., {et~al.} 2018, \mnras, 477,
  1206, \dodoi{10.1093/mnras/sty618}

\bibitem[{{Nelson} {et~al.}(2018){Nelson}, {Pillepich}, {Springel},
  {Weinberger}, {Hernquist}, {Pakmor}, {Genel}, {Torrey}, {Vogelsberger},
  {Kauffmann}, {Marinacci}, \& {Naiman}}]{TNG3}
{Nelson}, D., {Pillepich}, A., {Springel}, V., {et~al.} 2018, \mnras, 475, 624,
  \dodoi{10.1093/mnras/stx3040}

\bibitem[{{Patton} {et~al.}(2000){Patton}, {Carlberg}, {Marzke}, {Pritchet},
  {da Costa}, \& {Pellegrini}}]{Patton2000}
{Patton}, D.~R., {Carlberg}, R.~G., {Marzke}, R.~O., {et~al.} 2000, \apj, 536,
  153, \dodoi{10.1086/308907}

\bibitem[{{Patton} {et~al.}(2011){Patton}, {Ellison}, {Simard}, {McConnachie},
  \& {Mendel}}]{Patton2011}
{Patton}, D.~R., {Ellison}, S.~L., {Simard}, L., {McConnachie}, A.~W., \&
  {Mendel}, J.~T. 2011, \mnras, 412, 591,
  \dodoi{10.1111/j.1365-2966.2010.17932.x}

\bibitem[{{Patton} {et~al.}(2024){Patton}, {Faria}, {Hani}, {Torrey},
  {Ellison}, {Thakur}, \& {Westlake}}]{Patton2024}
{Patton}, D.~R., {Faria}, L., {Hani}, M.~H., {et~al.} 2024, \mnras, 529, 1493,
  \dodoi{10.1093/mnras/stae608}

\bibitem[{{Pearson} {et~al.}(2016){Pearson}, {Besla}, {Putman}, {Lutz},
  {Fern{\'a}ndez}, {Stierwalt}, {Patton}, {Kim}, {Kallivayalil}, {Johnson}, \&
  {Sung}}]{Pearson2016}
{Pearson}, S., {Besla}, G., {Putman}, M.~E., {et~al.} 2016, \mnras, 459, 1827,
  \dodoi{10.1093/mnras/stw757}

\bibitem[{{Petersen} \& {Pe{\~n}arrubia}(2021)}]{Petersen2021}
{Petersen}, M.~S., \& {Pe{\~n}arrubia}, J. 2021, Nature Astronomy, 5, 251,
  \dodoi{10.1038/s41550-020-01254-3}

\bibitem[{{Pillepich} {et~al.}(2018){Pillepich}, {Nelson}, {Hernquist},
  {Springel}, {Pakmor}, {Torrey}, {Weinberger}, {Genel}, {Naiman}, {Marinacci},
  \& {Vogelsberger}}]{TNG5}
{Pillepich}, A., {Nelson}, D., {Hernquist}, L., {et~al.} 2018, \mnras, 475,
  648, \dodoi{10.1093/mnras/stx3112}

\bibitem[{{Planck Collaboration} {et~al.}(2016){Planck Collaboration}, {Ade},
  {Aghanim}, {Arnaud}, {Ashdown}, {Aumont}, {Baccigalupi}, {Banday},
  {Barreiro}, {Bartlett}, {Bartolo}, {Battaner}, {Battye}, {Benabed},
  {Beno{\^\i}t}, {Benoit-L{\'e}vy}, {Bernard}, {Bersanelli}, {Bielewicz},
  {Bock}, {Bonaldi}, {Bonavera}, {Bond}, {Borrill}, {Bouchet}, {Boulanger},
  {Bucher}, {Burigana}, {Butler}, {Calabrese}, {Cardoso}, {Catalano},
  {Challinor}, {Chamballu}, {Chary}, {Chiang}, {Chluba}, {Christensen},
  {Church}, {Clements}, {Colombi}, {Colombo}, {Combet}, {Coulais}, {Crill},
  {Curto}, {Cuttaia}, {Danese}, {Davies}, {Davis}, {de Bernardis}, {de Rosa},
  {de Zotti}, {Delabrouille}, {D{\'e}sert}, {Di Valentino}, {Dickinson},
  {Diego}, {Dolag}, {Dole}, {Donzelli}, {Dor{\'e}}, {Douspis}, {Ducout},
  {Dunkley}, {Dupac}, {Efstathiou}, {Elsner}, {En{\ss}lin}, {Eriksen},
  {Farhang}, {Fergusson}, {Finelli}, {Forni}, {Frailis}, {Fraisse},
  {Franceschi}, {Frejsel}, {Galeotta}, {Galli}, {Ganga}, {Gauthier}, {Gerbino},
  {Ghosh}, {Giard}, {Giraud-H{\'e}raud}, {Giusarma}, {Gjerl{\o}w},
  {Gonz{\'a}lez-Nuevo}, {G{\'o}rski}, {Gratton}, {Gregorio}, {Gruppuso},
  {Gudmundsson}, {Hamann}, {Hansen}, {Hanson}, {Harrison}, {Helou},
  {Henrot-Versill{\'e}}, {Hern{\'a}ndez-Monteagudo}, {Herranz}, {Hildebrandt},
  {Hivon}, {Hobson}, {Holmes}, {Hornstrup}, {Hovest}, {Huang}, {Huffenberger},
  {Hurier}, {Jaffe}, {Jaffe}, {Jones}, {Juvela}, {Keih{\"a}nen}, {Keskitalo},
  {Kisner}, {Kneissl}, {Knoche}, {Knox}, {Kunz}, {Kurki-Suonio}, {Lagache},
  {L{\"a}hteenm{\"a}ki}, {Lamarre}, {Lasenby}, {Lattanzi}, {Lawrence}, {Leahy},
  {Leonardi}, {Lesgourgues}, {Levrier}, {Lewis}, {Liguori}, {Lilje},
  {Linden-V{\o}rnle}, {L{\'o}pez-Caniego}, {Lubin}, {Mac{\'\i}as-P{\'e}rez},
  {Maggio}, {Maino}, {Mandolesi}, {Mangilli}, {Marchini}, {Maris}, {Martin},
  {Martinelli}, {Mart{\'\i}nez-Gonz{\'a}lez}, {Masi}, {Matarrese}, {McGehee},
  {Meinhold}, {Melchiorri}, {Melin}, {Mendes}, {Mennella}, {Migliaccio},
  {Millea}, {Mitra}, {Miville-Desch{\^e}nes}, {Moneti}, {Montier}, {Morgante},
  {Mortlock}, {Moss}, {Munshi}, {Murphy}, {Naselsky}, {Nati}, {Natoli},
  {Netterfield}, {N{\o}rgaard-Nielsen}, {Noviello}, {Novikov}, {Novikov},
  {Oxborrow}, {Paci}, {Pagano}, {Pajot}, {Paladini}, {Paoletti}, {Partridge},
  {Pasian}, {Patanchon}, {Pearson}, {Perdereau}, {Perotto}, {Perrotta},
  {Pettorino}, {Piacentini}, {Piat}, {Pierpaoli}, {Pietrobon}, {Plaszczynski},
  {Pointecouteau}, {Polenta}, {Popa}, {Pratt}, {Pr{\'e}zeau}, {Prunet},
  {Puget}, {Rachen}, {Reach}, {Rebolo}, {Reinecke}, {Remazeilles}, {Renault},
  {Renzi}, {Ristorcelli}, {Rocha}, {Rosset}, {Rossetti}, {Roudier},
  {Rouill{\'e} d'Orfeuil}, {Rowan-Robinson}, {Rubi{\~n}o-Mart{\'\i}n},
  {Rusholme}, {Said}, {Salvatelli}, {Salvati}, {Sandri}, {Santos},
  {Savelainen}, {Savini}, {Scott}, {Seiffert}, {Serra}, {Shellard}, {Spencer},
  {Spinelli}, {Stolyarov}, {Stompor}, {Sudiwala}, {Sunyaev}, {Sutton},
  {Suur-Uski}, {Sygnet}, {Tauber}, {Terenzi}, {Toffolatti}, {Tomasi},
  {Tristram}, {Trombetti}, {Tucci}, {Tuovinen}, {T{\"u}rler}, {Umana},
  {Valenziano}, {Valiviita}, {Van Tent}, {Vielva}, {Villa}, {Wade}, {Wandelt},
  {Wehus}, {White}, {White}, {Wilkinson}, {Yvon}, {Zacchei}, \&
  {Zonca}}]{Planck2015}
{Planck Collaboration}, {Ade}, P.~A.~R., {Aghanim}, N., {et~al.} 2016, \aap,
  594, A13, \dodoi{10.1051/0004-6361/201525830}

\bibitem[{{Privon} {et~al.}(2017){Privon}, {Stierwalt}, {Patton}, {Besla},
  {Pearson}, {Putman}, {Johnson}, {Kallivayalil}, {Liss}, \&
  {Titans}}]{Privon2017}
{Privon}, G.~C., {Stierwalt}, S., {Patton}, D.~R., {et~al.} 2017, \apj, 846,
  74, \dodoi{10.3847/1538-4357/aa8560}

\bibitem[{{Robertson} {et~al.}(2019{\natexlab{a}}){Robertson}, {Dickinson},
  {Ferguson}, {Finkelstein}, {Furlanetto}, {Dayal}, {Greene}, {Hutter},
  {Madau}, {Marrone}, {Rhoades}, {Rhodes}, {Shapley}, {Stark}, {Wechsler}, \&
  {Zackrisson}}]{Robertson2019a}
{Robertson}, B., {Dickinson}, M., {Ferguson}, H.~C., {et~al.}
  2019{\natexlab{a}}, \baas, 51, 30

\bibitem[{{Robertson} {et~al.}(2019{\natexlab{b}}){Robertson}, {Banerji},
  {Brough}, {Davies}, {Ferguson}, {Hausen}, {Kaviraj}, {Newman}, {Schmidt},
  {Tyson}, \& {Wechsler}}]{Robertson2019b}
{Robertson}, B.~E., {Banerji}, M., {Brough}, S., {et~al.} 2019{\natexlab{b}},
  Nature Reviews Physics, 1, 450, \dodoi{10.1038/s42254-019-0067-x}

\bibitem[{{Rodriguez-Gomez} {et~al.}(2015){Rodriguez-Gomez}, {Genel},
  {Vogelsberger}, {Sijacki}, {Pillepich}, {Sales}, {Torrey}, {Snyder},
  {Nelson}, {Springel}, {Ma}, \& {Hernquist}}]{RG2015}
{Rodriguez-Gomez}, V., {Genel}, S., {Vogelsberger}, M., {et~al.} 2015, \mnras,
  449, 49, \dodoi{10.1093/mnras/stv264}

\bibitem[{{Snyder} {et~al.}(2017){Snyder}, {Lotz}, {Rodriguez-Gomez},
  {Guimar{\~a}es}, {Torrey}, \& {Hernquist}}]{Snyder2017}
{Snyder}, G.~F., {Lotz}, J.~M., {Rodriguez-Gomez}, V., {et~al.} 2017, \mnras,
  468, 207, \dodoi{10.1093/mnras/stx487}

\bibitem[{{Snyder} {et~al.}(2023){Snyder}, {Pe{\~n}a}, {Yung}, {Rose},
  {Kartaltepe}, \& {Ferguson}}]{Snyder2023}
{Snyder}, G.~F., {Pe{\~n}a}, T., {Yung}, L.~Y.~A., {et~al.} 2023, \mnras, 518,
  6318, \dodoi{10.1093/mnras/stac3397}

\bibitem[{{Spergel} {et~al.}(2015){Spergel}, {Gehrels}, {Baltay}, {Bennett},
  {Breckinridge}, {Donahue}, {Dressler}, {Gaudi}, {Greene}, {Guyon}, {Hirata},
  {Kalirai}, {Kasdin}, {Macintosh}, {Moos}, {Perlmutter}, {Postman},
  {Rauscher}, {Rhodes}, {Wang}, {Weinberg}, {Benford}, {Hudson}, {Jeong},
  {Mellier}, {Traub}, {Yamada}, {Capak}, {Colbert}, {Masters}, {Penny},
  {Savransky}, {Stern}, {Zimmerman}, {Barry}, {Bartusek}, {Carpenter}, {Cheng},
  {Content}, {Dekens}, {Demers}, {Grady}, {Jackson}, {Kuan}, {Kruk}, {Melton},
  {Nemati}, {Parvin}, {Poberezhskiy}, {Peddie}, {Ruffa}, {Wallace}, {Whipple},
  {Wollack}, \& {Zhao}}]{Spergel2015}
{Spergel}, D., {Gehrels}, N., {Baltay}, C., {et~al.} 2015, arXiv e-prints,
  arXiv:1503.03757, \dodoi{10.48550/arXiv.1503.03757}

\bibitem[{{Springel} {et~al.}(2001){Springel}, {Yoshida}, \&
  {White}}]{Springel2001b}
{Springel}, V., {Yoshida}, N., \& {White}, S.~D.~M. 2001, \na, 6, 79,
  \dodoi{10.1016/s1384-1076(01)00042-2}

\bibitem[{{Springel} {et~al.}(2018){Springel}, {Pakmor}, {Pillepich},
  {Weinberger}, {Nelson}, {Hernquist}, {Vogelsberger}, {Genel}, {Torrey},
  {Marinacci}, \& {Naiman}}]{TNG1}
{Springel}, V., {Pakmor}, R., {Pillepich}, A., {et~al.} 2018, \mnras, 475, 676,
  \dodoi{10.1093/mnras/stx3304}

\bibitem[{{Stewart} {et~al.}(2009){Stewart}, {Bullock}, {Barton}, \&
  {Wechsler}}]{Stewart2009}
{Stewart}, K.~R., {Bullock}, J.~S., {Barton}, E.~J., \& {Wechsler}, R.~H. 2009,
  \apj, 702, 1005, \dodoi{10.1088/0004-637x/702/2/1005}

\bibitem[{{Stierwalt} {et~al.}(2015){Stierwalt}, {Besla}, {Patton}, {Johnson},
  {Kallivayalil}, {Putman}, {Privon}, \& {Ross}}]{Stierwalt2015}
{Stierwalt}, S., {Besla}, G., {Patton}, D., {et~al.} 2015, \apj, 805, 2,
  \dodoi{10.1088/0004-637x/805/1/2}

\bibitem[{{Varma} {et~al.}(2022){Varma}, {Huertas-Company}, {Pillepich},
  {Nelson}, {Rodriguez-Gomez}, {Dekel}, {Faber}, {Iglesias-Navarro}, {Koo}, \&
  {Primack}}]{Varma2022}
{Varma}, S., {Huertas-Company}, M., {Pillepich}, A., {et~al.} 2022, \mnras,
  509, 2654, \dodoi{10.1093/mnras/stab3149}

\bibitem[{{Ventou} {et~al.}(2019){Ventou}, {Contini}, {Bouch{\'e}}, {Epinat},
  {Brinchmann}, {Inami}, {Richard}, {Schroetter}, {Soucail}, {Steinmetz}, \&
  {Weilbacher}}]{Ventou2019}
{Ventou}, E., {Contini}, T., {Bouch{\'e}}, N., {et~al.} 2019, \aap, 631, A87,
  \dodoi{10.1051/0004-6361/201935597}

\bibitem[{{Wuyts} {et~al.}(2012){Wuyts}, {F{\"o}rster Schreiber}, {Genzel},
  {Guo}, {Barro}, {Bell}, {Dekel}, {Faber}, {Ferguson}, {Giavalisco}, {Grogin},
  {Hathi}, {Huang}, {Kocevski}, {Koekemoer}, {Koo}, {Lotz}, {Lutz}, {McGrath},
  {Newman}, {Rosario}, {Saintonge}, {Tacconi}, {Weiner}, \& {van der
  Wel}}]{Wuyts2012}
{Wuyts}, S., {F{\"o}rster Schreiber}, N.~M., {Genzel}, R., {et~al.} 2012, \apj,
  753, 114, \dodoi{10.1088/0004-637X/753/2/114}

\bibitem[{{Wuyts} {et~al.}(2013){Wuyts}, {F{\"o}rster Schreiber}, {Nelson},
  {van Dokkum}, {Brammer}, {Chang}, {Faber}, {Ferguson}, {Franx}, {Fumagalli},
  {Genzel}, {Grogin}, {Kocevski}, {Koekemoer}, {Lundgren}, {Lutz}, {McGrath},
  {Momcheva}, {Rosario}, {Skelton}, {Tacconi}, {van der Wel}, \&
  {Whitaker}}]{Wuyts2013}
{Wuyts}, S., {F{\"o}rster Schreiber}, N.~M., {Nelson}, E.~J., {et~al.} 2013,
  \apj, 779, 135, \dodoi{10.1088/0004-637x/779/2/135}

\end{thebibliography}
